\documentclass[11pt]{article}
\usepackage{amsmath, amssymb, graphics}

\usepackage[nosort]{cite}
\newcommand{\mathsym}[1]{{}}
\usepackage{verbatim}
\usepackage{amsfonts,euscript,amssymb,stmaryrd,braket}
\usepackage{graphics,tikz}
\usetikzlibrary{shapes.misc,arrows,decorations.markings,patterns,calc,shapes.geometric}
\usepackage{caption}
\usepackage{subcaption}
\usepackage{slashed}
\definecolor{hyperref}{RGB}{026,028,185}
\usepackage[bookmarks=true,colorlinks=true,linkcolor=hyperref,citecolor=hyperref,urlcolor=hyperref,bookmarksnumbered]{hyperref}
 \usepackage{booktabs}
 \usepackage{multirow}
\usepackage{tabularx}
\usepackage{longtable}
\usepackage{bbold,bbding}
\usepackage{amsthm} 
\usepackage{esint}
\newcommand{\bal}{\begin{equation}\begin{aligned}}
\newcommand{\eal}{\end{aligned} \end{equation}}
\def\id{\protect{{1 \kern-.28em {\rm l}}}}

\makeatletter
\renewcommand\section{\@startsection {section}{1}{\z@}%
                                   {-3.5ex \@plus -1ex \@minus -.2ex}%
                                   {2.3ex \@plus.2ex}%
                                   {\normalfont\large\bfseries}}
\renewcommand\subsection{\@startsection{subsection}{2}{\z@}%
                                   {-3.25ex\@plus -1ex \@minus -.2ex}%
                                   {1.5ex \@plus .2ex}%
                                   {\normalfont\normalsize\bfseries}}

\makeatother

\numberwithin{equation}{section}

\usepackage[utf8]{inputenc}
\usepackage{epsfig}
\usepackage{graphicx}
\usepackage[multiple]{footmisc}
\usepackage{amssymb,amsmath}
\usepackage{fancybox,framed,tikz}
\usetikzlibrary{decorations.pathmorphing,patterns}
\usepackage{dsfont}
\usepackage{mathtools}
\usepackage{braket}
\usepackage{slashed}
\usepackage{rotating}
\usepackage{bbold,amsfonts}
\usepackage{multirow}
\usepackage{verbatim}
\usepackage{upgreek}
\usepackage{pgfplots}
\pgfplotsset{compat=1.17}

\tikzset{cross/.style={cross out, draw=black, minimum size=2*(#1-\pgflinewidth), inner sep=0pt, outer sep=0pt},
cross/.default={1pt}}

\newcommand{\be}{\begin{equation}}
\newcommand{\ee}{\end{equation}}

\newcommand{\G}{\Gamma}

\newcommand{\D}{\Delta}

\newcommand{\f}{\phi}

\newcommand{\Tr}{\textup{Tr}}

\usepackage{color}

\definecolor{mypink1}{rgb}{0.958, 0.188, 0.478}

\newcommand{\ba}{\begin{eqnarray}}
\newcommand{\ea}{\end{eqnarray}}

\hypersetup{}

\tikzset{Witten diagram/.style={execute at begin picture={%
\draw[blue ,fill=blue!05] circle[radius=\pgfkeysvalueof{/tikz/Witten/radius}];
\path node (X){\phantom{X}};
},baseline={(X.base)}},vertex/.style={circle,fill,inner sep=1.414pt,node
contents={}},
Witten/.cd,radius/.initial=1.414cm}

\begin{document}
\renewcommand{\thefootnote}{\arabic{footnote}}

\overfullrule=0pt
\parskip=2pt
\parindent=12pt
\headheight=0in \headsep=0in \topmargin=0in \oddsidemargin=0in

\begin{center}
\vspace{1.2cm}
{\Large\bf \mathversion{bold}
{Defects in the long-range O(N) model}\\
}
 
\author{ABC\thanks{XYZ} \and DEF\thanks{UVW} \and GHI\thanks{XYZ}}
 \vspace{0.8cm} {
 Lorenzo~Bianchi$^{a,b}$ \footnote{\tt lorenzo.bianchi@unito.it}, Leonardo S. Cardinale$^{c}$ \footnote{{\tt leonardo.cardinale@ens.psl.eu}}, Elia de Sabbata$^{b,d}$ \footnote{{\tt eliadesabbata@gmail.com}}
 }
 \vskip  0.5cm

\small
{\em
$^{a}$  
Dipartimento di Fisica, Universit\`a di Torino and INFN - Sezione di Torino\\ Via P. Giuria 1, 10125 Torino, Italy\\    

$^{b}$  
I.N.F.N. - sezione di Torino,\\
Via P. Giuria 1, I-10125 Torino, Italy\\    

$^{c}$  
Département de Physique, Ecole Normale Supérieure -- PSL \\ 
24 Rue Lhomond, 75005 Paris, France\\

$^{d}$  
Dipartimento di Scienze e Innovazione Tecnologica, Universit\`a del
Piemonte Orientale \\
Via T. Michel 11, 15121 Alessandria, Italy \\

\vskip 0.02cm

}
\normalsize

\end{center}

\vspace{0.3cm}
\begin{abstract} 
We initiate the study of extended excitations in the long-range O(N) model. We focus on line and surface defects and we discuss the challenges of a naive generalization of the simplest defects in the short-range model. To face these challenges we propose three alternative realizations of defects in the long-range model. The first consists in introducing an additional parameter in the perturbative RG flow or, equivalently, treating the non-locality of the model as a perturbation of the local four-dimensional theory. The second is based on the introduction of non-local defect degrees of freedom coupled to the bulk and it provides some non-trivial defect CFTs also in the case of a free bulk, i.e.~for generalized free field theory. The third approach is based on a semiclassical construction of line defects. After finding a non-trivial classical field configuration we consider the fluctuation Lagrangian to obtain quantum corrections for the defect theory.
\end{abstract}

\newpage

%%%%%%%%%%%%%%%%%%%%%%%%%%%%%%%%%%%%%%%%%%%%%
%%%%%%%%%%%%%%%%%%%%%%%%%%%%%%%%%%%%%%%%%%%%%
\tableofcontents
 \newpage  

 \section{Introduction and discussion}
Recent developments in our understanding of Quantum Field Theories (QFTs) have profoundly transformed how we conceptualize and approach these theories. Symmetries and general consistency conditions, such as unitarity and locality, have turned out to be much more constraining than we had previously thought, and finding the best ways to impose these constraints has spurred a major line of research. These considerations have led to several important results in various areas, such as scattering amplitudes, cosmology and conformal field theory.  The application to the latter, commonly known as the conformal bootstrap, has yielded some unprecedented predictions for the critical exponents of statistical models (see \cite{Rychkov:2023wsd} for a review), paving the way for a fruitful exchange with the condensed matter community.
\medskip

While the assumption of locality is very natural in the context of QFTs, we know that realistic statistical models, such as spin lattices, might have long-range interactions, leading to a non-local description in the continuum limit. The Long-Range Ising (LRI) model, for instance, is a generalization of the Short-Range Ising (SRI) model where the interaction is weighted by a power of the distance between the nodes. The continuum description of this theory is a non-local quantum field theory, which displays an interesting phase diagram depending on the value of the exponent in the power law \cite{Fisher:1972zz,PhysRevB.15.4344,PhysRevB.8.281,aizenman_critical_1988}. In particular, for dimension $2<d<4$, there is an interesting region of the phase diagram where the fixed point is actually an interacting non-local conformal field theory. This phase space has been known since the seventies \cite{Fisher:1972zz,PhysRevB.15.4344,PhysRevB.8.281,aizenman_critical_1988} and its shape has been supported both by RG flow analyses \cite{Honkonen_1989,Honkonen_1990} and Monte Carlo simulations \cite{Picco:2012ak,Angelini_2014,PhysRevLett.89.025703,Zhao_2023}. More recently, the authors of \cite{Paulos:2015jfa} provided a formal proof of the conformal invariance of the model, while in \cite{Behan:2017dwr, Behan:2017emf} a novel UV description was proposed to study the LRI model at the crossover with the SRI. This revival of the model led to a wealth of new developments \cite{Slade:2016yer,Behan:2018hfx,Giombi:2019enr,Adelhardt:2020rkh,Benedetti:2020rrq,Chai:2021arp,Chakraborty:2021lwl,Chai:2021wac,Giombi:2022gjj,Benedetti:2023pbt,Benedetti:2024oif,Rong:2024vxo,Li:2024uac} including the application of bootstrap techniques \cite{Behan:2023ile} to four-point correlation functions.
\medskip

In this paper, we initiate the study of the LRI model in the presence of defects. Defects are important observables in QFTs and statistical mechanics \cite{Billo:2016cpy}. In particular, for critical systems, they can be used to model impurities, boundaries or domain walls. In the context of the SRI model and, more generally, for the short-range critical $O(N)$ model, a wide range of defects is currently known \cite{Vojta_2000,sachdev_quantum_1999,Sachdev_2001,Sachdev_2003,Deng:2005dh,Liendo:2012hy,Billo:2013jda,Bissi:2018mcq,ParisenToldin:2020gpb,Metlitski:2020cqy,Liu_2021,Toldin:2021kun,Cuomo:2021kfm,Padayasi:2021sik,Cuomo:2022xgw,Nishioka:2022odm,Rodriguez-Gomez:2022gbz,Gimenez-Grau:2022czc,Bianchi:2022sbz,Gimenez-Grau:2022ebb,Rodriguez-Gomez:2022xwm,Rodriguez-Gomez:2022gif,Nishioka:2022qmj,Krishnan:2023cff,Trepanier:2023tvb,Raviv-Moshe:2023yvq,Giombi:2023dqs,Pannell:2023pwz, Bianchi:2023gkk,Pannell:2024hbu,Shachar:2024cwk,deSabbata:2024xwn}. Here we take the first steps towards the generalization of these constructions to the case of the long-range O(N) model. There are different methods to approach the long-range fixed point in the literature, including large $N$, large charge and $\varepsilon$ expansions. In the latter case, one of the attractive features of long-range models is that the presence of an additional parameter $\sigma$ for the strength of the interaction, which decays as $|x|^{d+\sigma}$, allows us to perform an $\varepsilon$-expansion at fixed dimension $d$, where $\varepsilon=2\sigma-d$ parametrizes the distance from the crossover $\sigma=\frac{d}{2}$ with the Gaussian theory. The goal of this work is to introduce defects in the UV theory and look for the existence of non-trivial defect CFTs in the IR through an $\varepsilon$-expansion analysis of the RG flow.

\subsection*{Summary of the results}

We start our analysis from line and surface defects, which are realized, in the short-range model, by integrating classically marginal operators on the defect. The simplest examples are the localized magnetic field, where a single scalar field is integrated along a line, and surface defects involving the integration of a quadratic combination of fields. The general idea of these constructions is to analyze the behavior of the defect coupling in the $\varepsilon$-expansion and look for non-trivial perturbative fixed points. A naive generalization of these constructions turns out to work only close to four dimensions, where the integrated operators are classically marginal. As we mentioned, one of the important features of the long-range models is the possibility to work at fixed $d$ expanding around the crossover. Therefore, some non-trivial generalization is required to study defects in this regime. This is one of the goals of the present work. A more technical overview of our approach is given in section \ref{classification}. Here we shortly summarize our findings:
\begin{itemize}
 \item \emph{Non-local defects} (Section \ref{non-local}). We construct a whole new class of defects by introducing non-local scalar degrees of freedom on the defect and coupling them with the bulk scalars. We study some general restrictions on the form of such couplings and then we analyse the existence of some non-trivial fixed point in the simplest cases. One important consequence of this approach is that we will be able to construct non-trivial defects in generalized free field (GFF) theory.
 \item \emph{Defects close to four dimensions} (Section \ref{epsexp}). We generalize the construction of line and surface defects for the short-range Ising model using the ordinary $\varepsilon$-expansion, but introducing an additional paremeter that allows us to explore the long-range IR fixed points. This is equivalent to interpreting the non-locality of the model as a perturbation of the local four-dimensional theory.
 \item \emph{Semiclassical defects} (Section \ref{semiclassics}). To describe the localized magnetic field away from four dimensions, we develop a semiclassical approach. We look for non-trivial classical solutions for the field, whose form is consistent with the requirements of defect conformal field theories (in particular this solution is singular on the defect profile). Perturbation theory around this saddle provides a complicated fluctuation Lagrangian, which can be used to compute various observables in the defect CFT.
\end{itemize}

\subsection*{Outlook}
This paper is the first attempt to introduce impurities in the long-range O(N) model and, as such, it opens up a variety of new interesting future directions. The first comment is that we used only one of the techniques that are available in these models, i.e.~the $\varepsilon$-expansion. It would be interesting to analyze our constructions with other methods, such as large N or large charge expansions.

Non-perturbative approaches, such as Monte Carlo simulations or numerical bootstrap would also be extremely useful. For the former, it would be interesting to understand how to realize on the lattice the non-local defects we construct here. For the latter, the most natural question is whether one could find interesting constraints in the space of defects, even starting from the case of generalized free field theories, where we found a new class of interesting defects. In this respect, an active research direction focuses on the study of conformal defects in free theories \cite{Lauria:2020emq,Bianchi:2021snj,Chalabi:2022qit,DiPietro:2023gzi,Bashmakov:2024suh} and our work goes in the direction of understanding what happens if one relaxes the condition of locality.

Pushing our perturbative analysis to higher orders is an obvious and natural development and it would be interesting to combine this with the analytic bootstrap approach, along the lines of what has been done for the short-range model \cite{Gimenez-Grau:2022czc,Bianchi:2022sbz,Gimenez-Grau:2022ebb,Bianchi:2023gkk}. The most natural observables for this analysis would be the defect four-point function and the bulk two-point function, where the crossing equations might help to recover the perturbative results by symmetry considerations, without resorting to Feynman diagrams.

The part of the LRI phase space that we explore in this paper is formed by the region around $d=4$ and that around the crossover $\sigma=d/2$ with the Gaussian theory. Nevertheless, there is also another interesting region one could explore that is the ``upper'' crossover, drawing the boundary between the LRI and SRI phases. A novel perspective on how to approach this crossover has been put forward in \cite{Behan:2017emf} and it would be interesting to insert defects into that picture.

Finally, let us mention that the long-range perturbation theory has been recently used as a complementary way to extract information abuout the short-range model \cite{Rong:2024vxo}. This has the advantage that one can work at fixed space dimension, but, since perturbation theory is performed around the Gaussian crossover, obtaining information about the opposite crossover requires a resummation of the perturbative series. Still, it is an interesting approach and it would be nice to exploit it also in the case of defects. For instance, one natural question regards the fate of the non-local defects when we approach the SRI crossover. Do they produce new non-trivial defects in the short-range model?

 \section{The model and its defects}

\subsection{The long-range $O(N)$ model} \label{LRImodel}

This work focuses primarily on the long-range Ising (LRI) model and its $O(N)$ vector extension in dimensions $ 2 < d < 4 $. The long-range Ising model can be defined on a spin lattice through the following Hamiltonian:
\begin{equation}
H = - J \sum_{i \neq j} \frac{s_i s_j}{|i-j|^{d+\sigma}}\,,
\end{equation}
where we take $J > 0$ to stick to the ferromagnetic case. This model undergoes a second-order phase transition at a certain critical temperature $T = T_c$ (this has been proven in $d=1$ for some values of $\sigma$ in \cite{Dyson:1968up, frohlich1982phase, duminil2020long}). Following the standard Landau-Ginzburg approach, at $T = T_c$, it is possible to replace the discrete model with the following action of an interacting real massless scalar field
\begin{equation}
S = \frac{\mathcal{N}_\sigma}{2} \int d^dx d^dy \frac{\phi(x) \phi(y)}{|x-y|^{d+\sigma}} + \frac{\lambda_0}{4!} \int d^dx 	\, \phi(x)^4\,,
\label{LRI action}
\end{equation}
with a normalization constant $\mathcal{N}_\sigma$ fixed such that in momentum space the kinetic part of the action reads $\tfrac{1}{2} \int \tfrac{d^dp}{(2\pi)^p} \phi(-p) |p|^\sigma\phi(p)$. Such a normalization is given by

\begin{equation}
\mathcal{N}_{\sigma}=\frac{2^{\sigma}\Gamma\left(\frac{d+\sigma}{2}\right)}{\pi^\frac{d}{2}\Gamma\left(-\frac{\sigma}{2}\right)}\,.
\end{equation}
Henceforth, we will refer to this continuum action as the LRI model. The generalization to the long-range $O(N)$ model is straightforward. One can promote $\phi$ to an $O(N)$ vector field, $\phi^a$, where $a = 1, \dots, N$, and contract the indices in the natural way
\begin{equation}
S = \frac{\mathcal{N}_\sigma}{2} \int d^dx d^dy \frac{\phi_a(x) \phi_a(y)}{|x-y|^{d+\sigma}} + \frac{\lambda_0}{4!} \int d^dx \left(\phi_a(x)\phi_a(x)\right)^2 \,. \label{LRONaction} 
\end{equation}
\noindent
The case $N = 1$ reduces to the LRI model \eqref{LRI action}.

The actions \eqref{LRI action} and \eqref{LRONaction} are clearly non-local.
More precisely, an action is \emph{local} if it involves a single integral of the fields and their derivatives up to a finite order.
In contrast, actions that depend on an infinite number of field derivatives or involve more than one integral are usually referred to as \emph{non-local} (see \cite{Hernandez-Cuenca:2024pey} for a modern treatment of this subject).
The non-local kinetic term can also be rewritten using of the fractional Laplacian $\mathcal{L}_\sigma = (-\partial^2)^{\sigma/2}$, where  $\sigma$ is a real number.
This operator acts on plane waves as  $\mathcal{L}_\sigma e^{ipx} = |p|^\sigma e^{ipx}$, and in position space it is given by
\begin{equation}
\mathcal{L}_\sigma \phi(x) = \mathcal{N}_{\sigma} \int d^dy \frac{\phi(y)}{|x-y|^{d + \sigma}}\,.
\end{equation}
Using this notation, the kinetic term becomes
\begin{equation}\label{nonlocalkineticterm}
 S_\text{kin}=\frac{1}{2} \int d^d x \, \phi(x) \mathcal{L}_{\sigma} \phi(x)\,,
\end{equation}
which leads to the following classical scaling dimension
\begin{equation}\label{Deltaphi}
\Delta_\phi = \frac{d-\sigma}{2}\,.
\end{equation}
When $\sigma=2$ the fractional Laplacian reduces to the usual Laplacian and the theory becomes the more familiar quartic theory that describes the short-range Ising (SRI) model.

\begin{figure}[htbp]
    \centering
\includegraphics[width=0.8\linewidth]{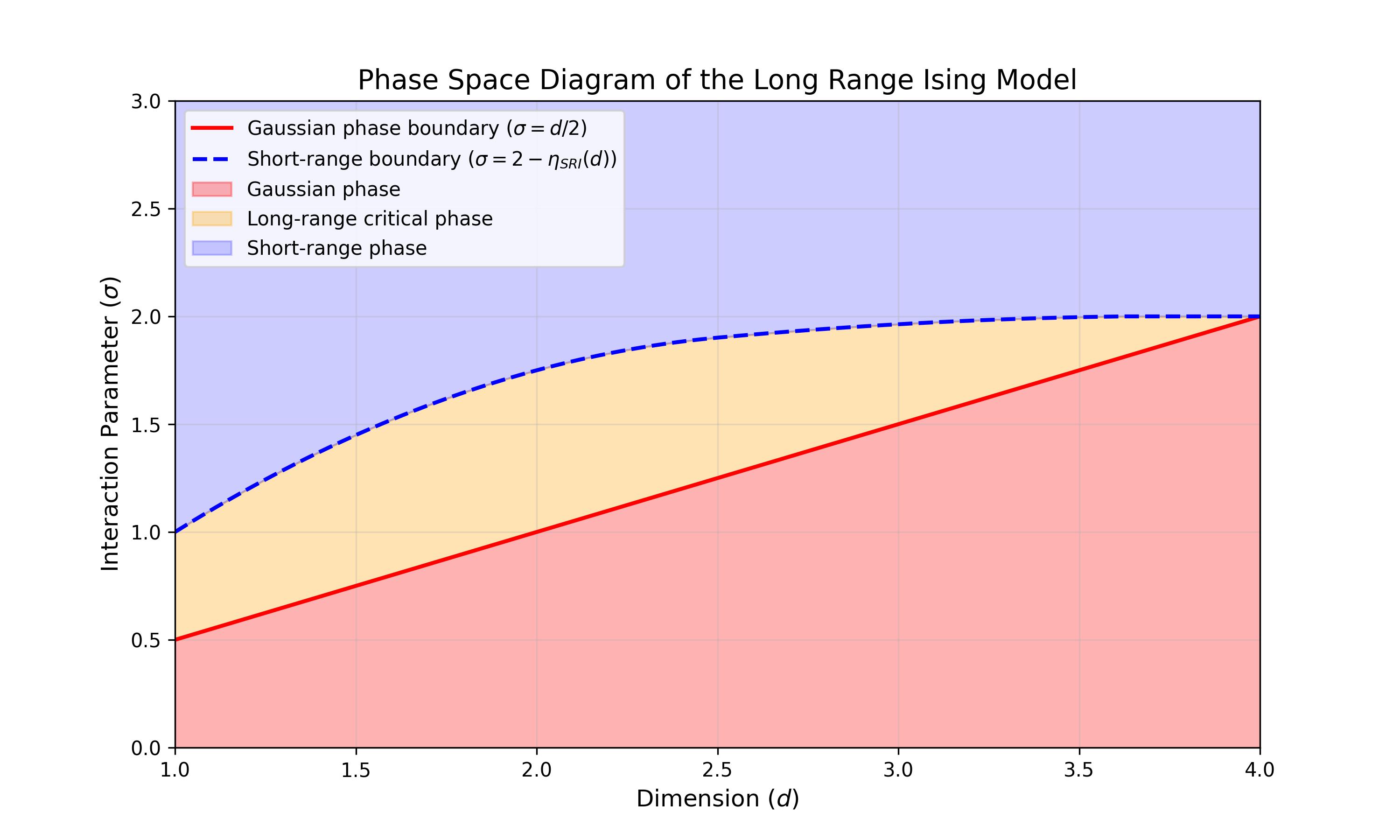}
\caption{Phase space diagram for the long-range Ising model as a function of the dimension $d$ and the parameter $\sigma$. The diagram shows the Gaussian phase ($\sigma \leq d/2$), the long-range critical phase ($d/2 < \sigma < 2 - \eta_{SRI}(d)$), and the short-range phase ($\sigma \geq 2 - \eta_{SRI}(d)$).}
    \label{phases}
\end{figure}

 The LRI has a rich phase space structure that depends on the value of $\sigma$  \cite{Paulos:2015jfa,Behan:2019lyd,Behan:2017emf}. This is illustrated in Figure \ref{phases}:

\begin{itemize}
  \item For $\sigma < d/2$, the quartic interaction is irrelevant, and the theory becomes free in the infrared (IR). In particular, it is a generalized free field (GFF) theory with $\Delta_{\phi}$ given in \eqref{Deltaphi}.
  \item For $\sigma > \sigma_{*} = d - 2 \Delta_{\phi}^{\text{SRI}}$, where $\Delta_{\phi}^{\text{SRI}}$ is the conformal dimension of $\phi$ in the short-range Ising model, the critical theory reduces to the short-range Ising model.
  \item For $d/2 < \sigma < \sigma_{*}$, the critical theory is non-trivial and non-Gaussian. Indeed, the quartic interaction is relevant and drives the theory towards an IR fixed point, known as the LRI fixed point (see Figure \ref{phases}).
\end{itemize}

Similar statements apply to the $O(N)$ generalization of the LRI.

There is another way to get to the action in equation \eqref{LRI action}, where one interprets the LRI as a defect of dimension $d$ embedded in a bulk space of (generically non-integer) dimension $d - \sigma + 2$. Specifically, if one considers a local massless field $\Phi(x, y)$ with action
\begin{equation}
S = \int d^dx \, d^{2-\sigma}y \left[(\partial_x \Phi)^2 + (\partial_y \Phi)^2\right]+ \int_{y = 0} d^dx \, \Phi^4\,,
\end{equation}
then integrating out the bulk field $\Phi(x, y \neq 0)$ leads to a theory for $\phi(x) = \Phi(x, 0)$, which is identical to the LRI theory. This method is sometimes referred to as the Caffarelli-Silvestre trick \cite{caffarelli2007extension}. In fact, this perspective is particularly useful for proving that the LRI exhibits enhanced conformal symmetry at the IR fixed point \cite{Paulos:2015jfa}.

The free propagator of the long-range $O(N)$ model can be easily derived
\begin{equation}
G_{ab}(x) = \frac{2^{d-\sigma} \Gamma\left(\frac{d-\sigma}{2}\right)}{(4\pi)^{\frac{d}{2}}\Gamma\left(\frac{\sigma}{2}\right)} \frac{\delta_{ab}}{|x|^{d-\sigma}} = \frac{ \mathcal{N}_\phi^{\,2}\delta_{ab}}{|x|^{2\Delta_\phi}}\,.
\label{propagator}
\end{equation}
To study the LRI fixed point perturbatively, the standard approach is to perform an $\varepsilon$-expansion near the crossover $\sigma=d/2$, as shown in Figure \ref{phases}. However, it is also useful to introduce a second type of expansion close to the corner $d=4$. For clarity and to set up the notation, we present both expansions here.
\begin{itemize}
 \item  Setting $ \sigma = \frac{d + \varepsilon}{2} $, the dimension of the field $ \phi $ in equation \eqref{Deltaphi} becomes $ \Delta_{\phi} = \frac{d - \varepsilon}{4} $, so that the $ \phi^4 $ interaction is weakly relevant, allowing for a perturbative expansion in $ \varepsilon $. An important and interesting feature of this expansion is that it can be performed at a fixed $ d $. The perturbative $ \beta $-function for the $ \lambda $ coupling has a non-trivial zero, corresponding to an infrared (IR) fixed point. The standard $ \varepsilon $-expansion procedure can then be used to compute observables at the perturbative fixed point, with the crucial difference that the field $ \phi $, being non-local, does not renormalize. This is the expansion that was first introduced in \cite{Fisher:1972zz}.

  \item Following the standard procedure, one can set $ d = 4 - \hat{\varepsilon} $, while introducing an additional parameter $ \kappa $ such that $ \sigma = 2 - \frac{(1 - \kappa) \hat{\varepsilon}}{2} $. This new parameter defines a specific direction when moving away from the corner $ d = 4 $ in Figure \ref{phases}. For $ \kappa = 1 $, we have $ \sigma = 2 $, and the action is local, corresponding to the ordinary short-range Ising (SRI) fixed point. The opposite limit is $ \kappa = 0 $, where $ \sigma = d/2 $, and one moves along the crossover to the Gaussian theories. All intermediate values $ 0 < \kappa < 1 $ correspond to the LRI fixed points. A similar expansion has been considered in \cite{honkonen1989crossover}.
\end{itemize}
A large $ N $ expansion is also possible \cite{Giombi:2019enr}, although it will not be considered in this work. The second expansion is particularly useful in Section \ref{epsexp}, where we consider the generalization of the localized magnetic field to the long-range model. Here, we focus on the first expansion and briefly review the renormalization procedure. 

As mentioned earlier, the kinetic term of the action is non-local, and since renormalization always introduces local counterterms, the field $ \phi $ does not renormalize. This implies that all divergences must be canceled by renormalizing the coupling $ \lambda $. We can express the renormalized coupling in terms of the bare coupling as $ \lambda_0 =\mu^\varepsilon Z_\lambda \lambda  $, where $ \mu $ is a mass scale. The action is then written as
\begin{equation}
S = \frac{\mathcal{N}_\sigma}{2} \int d^dx d^dy  \frac{\phi_a(x) \phi_a(y)}{|x-y|^{d+\sigma}} + \frac{Z_\lambda\lambda \mu^\varepsilon}{4!} \int d^dx \left(\phi_a(x)\phi_a(x)\right)^2\,,
\end{equation}
where $Z_\lambda$ ensures the cancellation of all poles in $\varepsilon$ in the correlators of $\phi$. At two-loop order, in the MS scheme, one gets
\begin{equation}\label{Zlambda}
\begin{split}
Z_\lambda & = 1 + \frac{\lambda (N+8)}{3(4\pi)^{\frac{d}{2}}\Gamma\!\left(\frac{d}{2}\right)\varepsilon}  + \frac{\lambda^2}{9 (4\pi)^d\Gamma\!\left(\frac{d}{2}\right)^2} \Bigg(\frac{N+8}{\varepsilon^2} \, + \\
& \hspace{3 cm}- \frac{(5N+22)\left(\psi\left(\frac{d}{2}\right) - 2 \psi\left(\frac{d}{4}\right) - \gamma_E\right)}{\varepsilon}\Bigg)+\text{O}(\lambda^3)\,.
\end{split}
\end{equation}
The beta function $\beta_{\lambda}=\mu \frac{\partial \lambda}{\partial{\mu}}$ is obtained by imposing $\mu \frac{d}{d\mu}\lambda_0=0$, and it is \cite{Fisher:1972zz, Giombi:2019enr}\, 
\begin{equation}
\beta_\lambda = - \varepsilon \lambda + \frac{(N+8)\lambda^2}{3(4 \pi)^{\frac{d}{2}}\Gamma\! \left(\frac{d}{2}\right)} - \frac{2(5N+22)(\psi\left(\frac{d}{2}\right) - 2 \psi\left(\frac{d}{4}\right) - \gamma_E)\lambda^3}{9(4\pi)^d\Gamma\!\left(\frac{d}{2}\right)^2}  + \text{O}(\lambda^4)\,.
\end{equation} 
This beta function has a non-trivial zero, corresponding to a perturbative fixed point
\begin{equation}
\frac{\lambda_\ast}{\Gamma\left(\frac{d}{2}\right)\left(4\pi\right)^\frac{d}{2}}=\frac{3}{N+8}\varepsilon+\frac{6\left(5N+22\right)\left(\psi\left(\frac{d}{2}\right)-2\psi\left(\frac{d}{4}\right)-\gamma_E\right)}{\left(N+8\right)^3}\varepsilon^2+\text{O}\left(\varepsilon^3\right)\,.
\end{equation}
The conformal invariance of this long-range $O(N)$ fixed point can be ascertained by extending the arguments of \cite{Paulos:2015jfa} for the LRI model.

\subsection{Existence and classification of some non-trivial defects}
\label{classification}
The goal of this work is the construction of non-trivial conformal defects in the non-local $O(N)$ model that we have just introduced. As a side result, we will also discuss the existence of non-trivial defects in the particular case $\lambda=0$, \emph{i.e.}~for GFF theories. Indeed, it has been shown that in integer dimensions less than four, free local theories do not admit any non-trivial defects \cite{Lauria:2020emq}, where trivial means Gaussian. We will show that, dropping the assumption of locality, we will be able to construct non-trivial conformal defects for GFF in three dimensions.

We begin by describing the most straightforward defects that can be constructed. The first type is obtained by integrating a power of one of the fields, $\phi^a$, along a line, thereby breaking $O(N)$ symmetry down to $O(N-1)$. This is analogous to the localized magnetic field, as discussed in \cite{Allais:2014fqa, Cuomo:2021kfm}.
We will explain why this construction does not lend itself easily to an expansion around $\sigma = d/2$ at fixed $d$, and introduce two alternative strategies to overcome this challenge, which we will refer to as the \emph{semiclassical} and \emph{non-local} defects.

Another simple defect is obtained by integrating the singlet $\phi_a \phi_a$ over a surface. The analogous defect for the local $O(N)$ model is discussed in \cite{Trepanier:2023tvb, Raviv-Moshe:2023yvq, Giombi:2023dqs}. Similar considerations apply in this case as well.

Of course, one could also consider other types of defects, such as magnetic impurities or monodromy defects. However, in this paper, we focus on the simplest constructions.

\subsubsection{Local defects}
As mentioned earlier, the simplest approach to constructing defects is to integrate integer powers of the fields $\phi_a$ over a $p$-dimensional subspace. Thus, we can consider the action
\begin{equation} S = S_\text{bulk} + h_0^{a_1 \dots a_n}\int d^p\tau \, \phi_{a_1}(\tau) \dots \phi_{a_n}(\tau)\,, \end{equation}
where $S_\text{bulk}$ is given by \eqref{LRONaction}, and $h_0^{a_1 \dots a_n}$ is a coupling constant. The usual strategy is to compute the beta function for the coupling $h$ and look for perturbative fixed points in the $\varepsilon$-expansion. For the expansion to be reliable, we need the coupling to be classically marginal, which requires $n\Delta_{\phi} = p$.
Since the dimension of $\phi$ is determined by the bulk theory, we have
\begin{equation} n = \frac{2p}{d - \sigma}\,.
\end{equation}
At the crossover point $\sigma = d/2$, we find $n = \frac{4p}{d}$, which, for $2 \leq d \leq 4$, is an integer only if $d = 2$ or $d = 4$. For this reason, this construction, unlike the case of the homogeneous long-range $O(N)$ model, forces us to consider the $\hat{\varepsilon}$-expansion around $d = 4$, and we cannot work at fixed $d$.

In Section \ref{non-local}, we will explore the $\hat{\varepsilon}$-expansion for the cases $p = n = 1$ (\emph{i.e.}~the long-range generalization of the localized magnetic field \cite{Allais:2014fqa, Cuomo:2021kfm}), and $p = n = 2$ (\emph{i.e.}~the generalization of the surface defects studied in \cite{Trepanier:2023tvb, Giombi:2023dqs}).

\subsubsection{Non-local defects}\label{subsubnonloc}

One of the main advantages of the non-local model is that it can be studied at fixed $d$. Therefore, we would like to find a defect construction that is suitable for the $\varepsilon$-expansion near the crossover line $\sigma = d/2$.
To achieve this, we introduce additional defect degrees of freedom with a non-local defect action, and couple them to the bulk via a local interaction.\,\footnote{
A similar construction can be found in \cite{Bashmakov:2024suh}, where free scalar theories are coupled to lower dimensional CFTs living on a defect.
}
For simplicity, let us consider the case $N = 1$. 
We introduce an additional bosonic field $\hat{\psi}$ on the defect and consider the following class of actions:
\begin{equation}\label{nonlocaldefectaction}
S = \int d^dx \left(\frac{1}{2} \phi \mathcal{L}_{\sigma} \phi + \frac{\lambda_0}{4!} \phi^4\right) +  \int d^p \tau \left( \frac{1}{2} \hat{\psi}\mathcal{L}_{\hat{\sigma}} \hat{\psi}  + \frac{g_0}{2} \phi^{a} \hat{\psi}^b \right)\,,
\end{equation}
where $\hat{\sigma} = p - 2 \hat{\Delta}_{\hat{\psi}}$, and $a$ and $b$ are integers. We can then tune the dimension $\hat{\Delta}_{\hat{\psi}}$ so that the coupling $g_0$ becomes classically marginal. This yields the condition
\begin{equation}
  a \Delta_{\phi}+ b \hat{\Delta}_{\hat{\psi}} \sim p\,,
\end{equation}
where the relation holds up to a small regulator that must be introduced to cure the divergences of the theory. Near the crossover $\sigma = d/2$, we find
\begin{equation}
\Delta_{\phi}\sim \frac{d}{4}\,, \qquad \hat{\Delta}_{\hat{\psi}} \sim  \frac{4p-ad}{4b}\,.
\end{equation}
Note that these dimensions will not receive quantum corrections as one moves away from the crossover, since both kinetic terms are non-local. To further constrain the allowed values of the exponents $a$ and $b$, we impose the unitarity condition
\begin{equation}
\hat{\Delta}_{\hat{\psi}} \geq \max \left(0,\frac{p-2}{2}\right)\,.
\label{scaling}
\end{equation}
The valid choices of parameters for a unitary theory are summarized below for $3 \leq d \leq 4$:
\begin{itemize}
\item \boldmath $p = 1$ \unboldmath: $a=1$, any $b$
\item \boldmath $p = 2$ \unboldmath: $a=1$ or $2$, any $b$
\end{itemize}
%The possible scenarios are summarized in Table \ref{allowed}.

%\begin{table}[htbp]
%\centering
%\includegraphics[scale=0.3]{Images/Unitary nonlocal defects}
%\caption{Configurations of parameters allowed for unitary non-local defects.}
%\label{allowed}
%\end{table}

Of course, having a classically marginal defect interaction is not sufficient to produce a non-trivial conformal defect in the IR. While the renormalization of the coupling $\lambda$ is unaffected by the presence of the defect, one must search for non-trivial fixed points of the coupling $g$ by computing its beta function and looking for a non-trivial zero. Section \ref{non-local} will be devoted to this analysis for specific cases with low values of $a$ and $b$. In particular, we will show that these defects are admissible even in the case $\lambda = 0$, \emph{i.e.}~for GFF theory in $d = 3$. 

Moreover, for $b = 1$ or $2$, the defect field $\hat{\psi}$ can be integrated out in the action \eqref{nonlocaldefectaction} (see appendix \ref{integratingoutpsi} for details). This leads to an effective action for $\phi$, providing an alternative approach to computing correlation functions of $\phi$.

\subsubsection{Semiclassical defects}

Another possible strategy for constructing conformal defects at fixed $d$ in the long-range $O(N)$ model is to investigate the IR fixed points of defect RG flows triggered by strongly relevant interactions. In general, this is a challenging problem, as the defect couplings are not perturbatively small in this regime. However, in certain cases, it is possible to study a strongly coupled defect in a weakly coupled bulk theory by expanding the action around a classical configuration that corresponds to a saddle point of the path integral.
In our case, the classical configuration for the field $\phi(x)$ must be consistent with the constraints imposed by the defect conformal symmetry. For example, in the case of an $O(N)$ symmetry-breaking line, we must have
\begin{equation}
\phi^a_{\text{cl}}(x) =  \delta_{a1} \frac{ \mathcal{N}_{\phi}\,a_\text{cl}}{|x_\perp|^{\Delta_{\text{cl}}}}\,.
\end{equation}
The constants $a_{\text{cl}}$ and $\Delta_\text{cl}$ can be determined by solving the classical equations of motion. The semiclassical expansion
\begin{equation}
\phi^a(x)= \phi^a_{\text{cl}}(x)+\delta \phi^a(x)\,,
\end{equation}
leads to an action for the fluctuation $\delta\phi$, which can be used to compute quantum corrections to observables in the defect field theory. 
This method has been used to study the $O(N)$ model in the presence of a boundary in various regimes \cite{ohno19841,McAvity:1995zd,Shpot:2019iwk,Metlitski:2020cqy}, and was also considered for surface defects in \cite{Giombi:2023dqs}.\footnote{See also \cite{Buhl-Mortensen:2016jqo,Kristjansen:2024map} for a similar approach in the context of defects in $\mathcal{N}=4$ super Yang-Mills.} However, some technical issues arise when one applies the same techniques to compute quantum corrections in defects with codimension $q=d-p\neq 1$.
We will discuss this construction in more detail in Section \ref{semiclassics}.

An alternative approach for studying strongly coupled defects in weakly coupled bulks, which we do not pursue in this work, is to use the methods introduced in \cite{deSabbata:2024xwn}. By considering a non-integer dimensional defect where the interaction is weakly relevant, it may be possible to compute observables and then extrapolate the results to integer values of the defect dimension.

 \section{Non-local defects}\label{non-local}

In this section we begin our analysis of the non-local defects introduced in Section \ref{subsubnonloc}. Our goal is to compute the beta function for the defect coupling $g$ and identify non-trivial perturbative zeros in the $\varepsilon$-expansion at fixed $d$. We focus primarily on the free bulk case $\lambda = 0$, and on the LRI model, \emph{i.e.}~$N = 1$. However, most of our results can be straightforwardly generalized to arbitrary $N$. Since the construction of these defects is carried out at a fixed dimension $d$, this analysis also serves as a study of the existence of non-trivial defects in GFF theory across different dimensions. In particular, we will show that GFF theory in three dimensions admits non-trivial conformal defects. Additionally, we will also show that introducing the bulk interaction, in most cases, does not affect the leading-order results for the beta function or the defect CFT data. 

Before examining specific values of the exponents $(a,b)$ in \eqref{nonlocaldefectaction}, we first establish some general results that hold for all RG flows considered in this section.
To begin, we need to specify how we regulate the divergences. We introduce a regulator $\varepsilon$ and impose that the defect interaction term in \eqref{nonlocaldefectaction} is weakly relevant \begin{equation} \label{deltaphihatregulator} a \Delta_\phi + b \hat{\Delta}_{\hat{\psi}} = p - \varepsilon\,. 
\end{equation} 
When the bulk theory is free, this choice is sufficient to ensure that all correlators remain finite.
In the case of an interacting bulk, we must also specify how to move away from the crossover at $\sigma = d/2$. As in Section \ref{LRImodel}, we set $\varepsilon = 2\sigma - d$, where $\varepsilon$ now regulates bulk integrals as well.
This choice is not unique; we could have inserted an arbitrary coefficient in front of the $\varepsilon$ in \eqref{deltaphihatregulator} (or, alternatively, use two independent regulators). However, as we will show, at leading order in $\varepsilon$ only defect integrals contribute, and it is straightforward to generalize the results introducing an additional coefficient.F
With this choice, the relation between the bare and the renormalized defect coupling is
\begin{equation}
 g_0= \mu^{\varepsilon} Z_g g(\mu) \,,
\end{equation}
and we can use this to compute the beta function.

To compute the beta function, one typically imposes that a given observable is finite by reabsorbing the divergences into the renormalization constants. A commonly used observable for this purpose is the bulk one-point function of the field $\phi$. However, in some models that we are considering, the perturbative computation of this one-point function involves tadpole integrals, which complicate the analysis.

Instead, we find it more convenient to look at the defect two-point function of a special composite operator, $\hat{\mathcal{O}}_0 = \hat{\phi}^{a-1} \hat{\psi}^b$, which appears in the equation of motion for $\phi$. Indeed, in the free bulk case we have
\begin{equation}\label{eomphinonloc}
\mathcal{L}_{\sigma} \phi(0,x_\perp) = - \frac{g_0}{2} a \, \hat{\mathcal{O}}_0(0) \delta^{(d-p)}(x_{\perp})\,.
\end{equation}
Since the bulk field $\phi$ does not renormalize (due to its non-local kinetic term), the left-hand side of \eqref{eomphinonloc} remains unchanged under renormalization. However, the right-hand side involves the renormalization of the coupling $g_0 = \mu^{\varepsilon} Z_g g$ and the wavefunction renormalization of the defect operator $\hat{\mathcal{O}}_0=Z_{\hat{\mathcal{O}}}\hat{\mathcal{O}}$. This implies that in the MS scheme, these two quantities are related by the condition $Z_g Z_{\hat{\mathcal{O}}} = 1$. Therefore, the renormalization of the coupling (and the corresponding beta function) can be derived from the wavefunction renormalization of $\hat{\mathcal{O}}$.
Indeed, at leading order, we expect the wavefunction renormalization to take the form
\begin{equation}
Z_{\hat{\mathcal{O}}} = 1 - \frac{\alpha g^n}{\varepsilon} + \text{O}\!\left(g^{n+1}\right)\,,
\end{equation}
where $n$ is an integer related to the values of $a$ and $b$, and $\alpha$ is a real number. The renormalization factor for the coupling is then given by
\begin{equation}
Z_g = 1 + \frac{\alpha g^n}{\varepsilon} + \text{O}\!\left(g^{n+1}\right)\,,
\end{equation}
and the beta function can be derived, as usual, by requiring that the bare coupling does not depend on the renormalization mass scale, leading to
\begin{equation}\label{betanonlocal}
\beta_g = - \varepsilon g + \alpha n g^{n+1} + \text{O}\!\left(g^{n+2}\right)\,.
\end{equation}
The non-trivial zero of the beta function is  $g_{*}^n = \varepsilon/(\alpha n) + \text{O}(\varepsilon^2)$. For odd values of $n$ this equation will always admit a real solution, while for even values of $n$ the crucial requirement is that $\alpha$ is positive. At this fixed point, the equation of motion \eqref{eomphinonloc} implies that the operator $\hat{\mathcal{O}}$ is protected with dimension
\begin{equation}
\hat{\Delta}_{\hat{\mathcal{O}}}= p - \Delta_\phi\,.
\end{equation}
This can also be checked by computing explicitly the anomalous dimension $\gamma_{\hat{\mathcal{O}}}$ which has the exact expression\,\footnote{A useful relation that can be used for this computation is $\beta_g=\frac{-\varepsilon g Z_g}{Z_g + g \partial_g Z_g}$.}
\begin{equation}
\gamma_{\hat{\mathcal{O}}} = \beta_g \frac{\partial \log Z_{\hat{\mathcal{O}}}}{\partial g} =\beta_g \frac{\partial \log Z^{-1}_g}{\partial g} = \varepsilon -\frac{\beta_g}{g}\,,
\end{equation}
and the second term vanishes at the fixed point.
Using also that $a \Delta_\phi +  b \hat{\Delta}_{\hat{\psi}} = p - \varepsilon$, we get $\hat{\Delta}_{\hat{\mathcal{O}}} = (a-1)\Delta_\phi + b \hat{\Delta}_{\hat{\psi}} + \gamma_{\hat{\mathcal{O}}}=p-\Delta_{\phi}$, as expected.
Therefore, we have
\begin{equation}\label{OOtwopoint}
\langle \hat{\mathcal{O}}(\tau_1) \hat{\mathcal{O}}(\tau_2) \rangle = \frac{\mathcal{N}_{\hat{\mathcal{O}}}^{\,2}}{|\tau_1-\tau_2|^{p-\Delta_\phi}}\,,
\end{equation}
where $\mathcal{N}_{\hat{\mathcal{O}}}$ is a normalization constant.

Furthermore, one can invert the equation of motion to express $\phi$ in terms of integrals of the operator $\hat{\mathcal{O}}$ on the defect. This allows to compute bulk correlators of $\phi$ by integrating defect correlators of $\hat{\mathcal{O}}$, without having to compute Feynman diagrams \cite{Bianchi:2023gkk}. Indeed, by inverting \eqref{eomphinonloc}, we find
\begin{equation}
\phi\left(x\right) = - a\, \mathcal{N}_\sigma  \frac{g_0}{2} \int d^p \tau \frac{\hat{\mathcal{O}}(\tau)}{\left(|x_{\perp}|^2 + |\tau|^2\right)^{\Delta_\phi}} + \phi_\text{free}(x) \,,
\label{inversion}
\end{equation}
where $\phi_\text{free}$ is a free field, which does not interact with the defect. This can be used to compute the one-point function of $\phi^2$
\begin{equation}\label{onepointphi2nonlocal}
\begin{aligned}
\langle \phi^2(x) \rangle &= g_{0}^2 \frac{a^2 \mathcal{N}_\sigma^{\,2}\,\mathcal{N}_{\hat{\mathcal{O}}}^{\,2}}{4}  \int \frac{d^p \tau_1 \, d^p \tau_2}{\left(|x_{\perp}|^2 + |\tau_1|^2\right)^{\Delta_\phi} |\tau_1 - \tau_2|^{2(p-\Delta_\phi)} \left(|x_{\perp}|^2 + |\tau_2|^2\right)^{\Delta_\phi}}=
\\ &=  g_{0}^2 \frac{a^2 \mathcal{N}_\sigma^{\,2}\,\mathcal{N}_{\hat{\mathcal{O}}}^{\,2}}{4}\, \frac{\pi^p \Gamma\left(\frac{p}{2}\right) \Gamma\left(\Delta_\phi - \frac{p}{2}\right)}{\Gamma\left(\Delta_\phi\right)(p-1)!} \frac{1}{|x_\perp|^{2\Delta_\phi}} = \frac{a_{\phi^2} \mathcal{N}_{\phi^2}}{|x_\perp|^{2\Delta_\phi}}\,,
\end{aligned}
\end{equation}
where $\mathcal{N}_{\phi^2}= 2\mathcal{N}_{\phi}^{\,2}$. The last step to compute a piece of defect CFT data is to evaluate this expression at the fixed point $g_{*}$. We will do this in a few specific cases below.

Similarly, we can compute the bulk-to-defect one-point functions of $\phi$ and $\hat{\mathcal{O}}$, as well as the bulk two-point function of $\phi$. For example, consider the correlator $\langle \phi(x) \phi(y) \rangle$. Following \cite{Bianchi:2023gkk}, we can exploit the residual conformal symmetry to set $x_\parallel = y_\parallel=0$ and $x_\perp=(z,\bar{z},0,\ldots)$, $y_\perp=(0,1,0,\ldots)$. It is also convenient to define a radial coordinate by $z \bar{z}=r$. Using \eqref{inversion} we get
\begin{equation}
\begin{split}
& \langle \phi(0,x_\perp) \phi(0,y_\perp) \rangle = \\
& \quad = g_{0}^2 \frac{a^2 \mathcal{N}_{2\Delta_\phi - d}^{\,2}\,\mathcal{N}_{\hat{\mathcal{O}}}^{\,2}}{4}\int \frac{d^p \tau_1 \, d^p \tau_2}{(1+|\tau_1|^2)^{\Delta_\phi}|\tau_1-\tau_2|^{2(p-\Delta_\phi)}(r^2+|\tau_2|^2)^{\Delta_\phi}}+ G(x-y)\,,
\end{split}
\end{equation}
where $G(x-y)$ is the free $\phi$ propagator. This integral can be computed exactly. By restoring the dependence on arbitrary $x$ and $y$ and evaluating at the fixed point, we get
\begin{equation}\label{twopointblock}
\begin{split}
&\langle \phi(x) \phi(y) \rangle = \frac{ \mathcal{N}_\phi^{\,2}\,F_{\phi\phi}(r)}{|x_\perp|^{\Delta_\phi}|y_\perp|^{\Delta_\phi}}\,, \\
&F_{\phi\phi}(r)= \xi^{-\Delta_\phi}+ g_*^2 \frac{a^2 \mathcal{N}_\sigma^{\,2}\,\mathcal{N}_{\hat{\mathcal{O}}}^{\,2}}{4 \mathcal{N}_\phi^{\,2}} \Bigg( \frac{\pi ^{p+2} r^{p-\Delta_\phi } \, _2F_1\!\left(\frac{p}{2},p-\Delta_\phi ;\frac{p}{2} - \Delta_\phi +1;r^2\right)}{\Gamma (\Delta_\phi )^2 \sin ^2\!\left( \pi  (\frac{p}{2}- \Delta_\phi )\right) \Gamma \! \left(\frac{p}{2}-\Delta_\phi +1\right)^2}\,+ \\
& \quad \quad -\frac{2 \pi ^{p+1} r^{\Delta_\phi } \, _2F_1\!\left(\frac{p}{2},\Delta_\phi ;1-\frac{p}{2}+\Delta_\phi ;r^2\right)}{\Gamma (\Delta_\phi ) (p-2 \Delta_\phi ) \sin \!\left( \pi  (\frac{p}{2}- \Delta_\phi )\right) \Gamma (p-\Delta_\phi )}\Bigg)\,,
\end{split}
\end{equation}
where $\xi=(x-y)^2/(|x_\perp||y_\perp|)$ is a conformal cross ratio.
From \eqref{twopointblock} we can immediately read off the spectrum of the exchanged operators in the defect channel operator product expansion (OPE). The only primary operators are $(\partial_\perp)^s \phi$ with $s$ integer and dimension $\hat{\Delta}_s=1+s-\varepsilon/2$, and the operator $\hat{\mathcal{O}}$ with dimension $\Delta_{\hat{\mathcal{O}}}=p-\Delta_\phi$.
one can also easily extract all the bulk-to-defect OPE coefficients for the operators exchanged in the defect channel, as well as the one-point function coefficients for the operators exchanged in the bulk channel, as showed in \cite{Bianchi:2023gkk}.

In the interacting bulk case, there are non-linear corrections to the left-hand side \eqref{eomphinonloc} and the above analysis does not apply.
However, in the models we consider, corrections due to the bulk coupling only appear at next-to-leading order. As a result, the above results remain valid perturbatively at leading order in $\varepsilon$.

We are now ready to compute the wavefunction renormalization $Z_{\hat{\mathcal{O}}}$ for some of the models discussed in Section \ref{subsubnonloc}
by looking at the two-point function $\langle \hat{\mathcal{O}}\hat{\mathcal{O}} \rangle $ using Feynman diagrams. 
This will enable us to compute the beta function and identify perturbative fixed points. Specifically, we will consider the cases  $(a,b)=(1,2)$,  $(a,b)=(2,1)$, and  $(a,b)=(2,2)$. 

Let us summarize the conventions used for the Feynman diagrams throughout this paper. A blue line always represents a defect: a straight blue line corresponds to a linear defect ( a line defect, a surface, etc.), while a blue circle represents a circular or (hyper)spherical defect. Solid black lines denote the field that is defined both in the bulk and on the defect (denoted by $\phi$), whereas dotted lines represent a field defined only on the defect (denoted by $\hat{\psi}$). 
We will use the following normalization constants
\begin{equation}
 \mathcal{N}_\phi^{\,2}= \frac{\G (\D_\f)}{2^{d-2\D_\f}\pi^\frac{d}{2}\,\G \!\left( \frac{d}{2}-\D_\f \right) }\,,\quad \quad \mathcal{N}_{\hat{\psi}}^{\,2}=\frac{\G (\hat{\Delta}_{\hat{\psi}})}{2^{p-2\hat{\Delta}_{\hat{\psi}}}\pi^\frac{p}{2}\,\G \!\left( \frac{p}{2}-\hat{\Delta}_{\hat{\psi}} \right) } \,. 
\end{equation}
Feynman rules are summarized in Table \ref{Feynmanrules}.
Additionally, we find it useful to define the following function \cite{Paulos:2015jfa}
\begin{equation}
w_\alpha^{(d)} = (4\pi)^{\frac{d}{2}} 2^{-\alpha} \frac{\Gamma\left(\frac{d-\alpha}{2}\right)}{\Gamma\left(\frac{\alpha}{2}\right)}\,.
\end{equation}

\begin{table}
\centering
\scalebox{0.8}{
\renewcommand{\arraystretch}{2} % Increases row height
\setlength{\tabcolsep}{12pt} % Adjusts column spacing
\begin{tabular}{|c|c|}
    \hline
    \textbf{Element} & \textbf{Rule} \\ \hline
    \begin{tikzpicture}[baseline={([yshift=-.5ex]current bounding box.center)}]
        % Internal propagator
        \draw[thick] (0, 0) -- (2, 0);
    \end{tikzpicture}
    & $\phi$ propagator: 
    \(
G(x-y) = \frac{\mathcal{N}_\phi^{\,2}}{|x-y|^{2\D_\phi}} \)  \\ \hline
     \begin{tikzpicture}[baseline={([yshift=-.5ex]current bounding box.center)}]
        % Internal propagator
        \draw[dotted, thick] (0, 0) -- (2, 0);
    \end{tikzpicture} &
    $\hat{\psi}$ propagator: 
    \(G(x-y) =  \frac{\mathcal{N}_{\hat{\psi}}^{\,2}}{|x-y|^{2\D_\psi}} \)
    \\ \hline
    \begin{tikzpicture}[baseline={([yshift=-.5ex]current bounding box.center)}]
        % Interaction vertex
        \draw[thick] (0, 0) -- (-0.5, 0.5);
        \draw[thick] (0, 0) -- (0.5, 0.5);
        \draw[thick] (0, 0) -- (-0.5, -0.5);
        \draw[thick] (0, 0) -- (0.5, -0.5);
        \fill (0, 0) circle (2pt);
    \end{tikzpicture}
    & bulk $\phi^4$ interaction
    \(-\frac{\lambda_0}{4!} \int d^d x\), where \(x\) is the interaction point
    \\ \hline
    \begin{tikzpicture}[baseline={([yshift=-.5ex]current bounding box.center)}]
        % Interaction vertex
        \draw[thick] (0, 0) -- (-0.5, 0.5);
        \draw[thick] (0, 0) -- (0.5, 0.5);
        \draw[thick,dotted] (0, 0) to[out=270,in=180] (-1, -0.5);
        \draw[thick,dotted] (0, 0) to[out=270,in=0] (1, -0.5);
        \draw[thick,double,blue] (-1,0)--(1,0);
        \draw[blue, fill=white]   (0,0) circle (2pt);
    \end{tikzpicture}
    & defect $\phi^a \hat{\psi}^b$ interaction
    \(-\frac{g_0}{2} \int d^p \tau\), where \(\tau\) is the interaction point on the defect \\ \hline
\end{tabular}
}
\caption{Feynman rules for the non-local defect.}
\label{Feynmanrules}
\end{table}

\subsection{$(a,b)=(1,2)$}

We start from $(a,b)=(1,2)$ in the free bulk case. The interaction term in \eqref{nonlocaldefectaction} becomes $ \frac{g_0}{2}  \int d^p \tau \phi \hat{\psi}^2$. The operator $\hat{\mathcal{O}}$ in this case is $\hat{\mathcal{O}} = \hat{\psi}^2$. The bare two-point function $\langle \hat{\psi}^2_0(0)\hat{\psi}_0^2(\tau)\rangle$  at tree level is given by the following diagram (dotted lines stand for free propagators of $\hat{\psi}$)
 \begin{equation}
 \begin{tikzpicture}[scale=0.4, baseline=-0.1cm]
	 \draw[double,thick,blue] (-3,0)--(3,0);
	  \draw[thick, dotted, black]    (-2,0) to[out=90,in=180] (0,2) to[out=0,in=90]  (2,0);
	   \draw[thick, dotted, black]    (-2,0) to[out=90,in=180] (0,1.3) to[out=0,in=90]  (2,0);
	  	 \draw[blue, fill=blue]   (-2,0) circle (4pt);
	  	  \draw[blue, fill=blue]   (2,0) circle (4pt);
	  	   \node[below] at (-2,0) {$\hat{\psi}^2$};
	  	   \node[below] at (2,0) {$\hat{\psi}^2$};
	\end{tikzpicture}
	 \, = \frac{2 (\mathcal{N}_{\hat{\psi}}^{\,2})^2}{|\tau|^{4\hat{\Delta}_{\hat{\psi}}}}\,.
 \end{equation}
 At one loop, there are three diagrams, but only one turns out to be divergent. Since we only need to compute $Z_{\hat{\psi}^2}$ , it suffices to consider this diagram 
\begin{equation}
\begin{aligned}
 \begin{tikzpicture}[scale=0.4, baseline=-0.1cm]
	 \draw[double,thick,blue] (-4,0)--(4,0);
	  \draw[thick, dotted, black]    (-3,0) to[out=90,in=180] (-1,1.5) to[out=0,in=90]  (1,0);
	  \draw[thick, dotted, black]    (-3,0) to[out=90,in=180] (-2,0.8) to[out=0,in=90]  (-1,0);
	   \draw[thick, black]    (-1,0) to[out=90,in=180] (0,0.8) to[out=0,in=90]  (1,0);
	   \draw[thick, dotted, black]    (1,0) to[out=270,in=180] (2,-0.6) to[out=0,in=270]  (3,0);
	  \draw[thick, dotted, black]    (-1,0) to[out=9270,in=180] (1,-1.2) to[out=0,in=270]  (3,0);
	  	 \draw[blue, fill=blue]   (-3,0) circle (4pt);
	  	  \draw[blue, fill=blue]   (3,0) circle (4pt);
	  	  \draw[blue, fill=white]   (-1,0) circle (4pt);
	  	  \draw[blue, fill=white]   (1,0) circle (4pt);
	  	  \node[below] at (-3,0) {$\hat{\psi}^2$};
	  	   \node[below] at (3.4,0) {$\hat{\psi}^2$};
	\end{tikzpicture} &= 2 \mathcal{N}_\phi^{\,2} \! \left(\mathcal{N}_{\hat{\psi}}^{\,2}\right)^{4} \!  g_0^2 \int \frac{d^p\tau_1 \, d^p\tau_2}{(|\tau_1|\,|\tau_2|\,|\tau - \tau_1|\,|\tau - \tau_2|)^{2\hat{\Delta}_{\hat{\psi}}}|\tau_2-\tau_1|^{2\Delta_\phi}}\,.
 \end{aligned}
\end{equation}
The divergent part of this diagram can be easily extracted by analyzing the limits $\tau_1, \tau_2 \to 0, \tau$
%\begin{equation}
%\frac{1}{|x|^{4\Delta_\psi}} \int d^py_1 d^py_2 \frac{1}{|y_1|^{2\Delta_\psi}} \frac{1}{|y_2|^{2\Delta_\psi}} \frac{1}{|y_2-y_1|^{2\Delta_\phi}} = \left(\frac{w_{2\Delta_\phi}^{(p)} w_{2\Delta_\psi}^{(p)}}{w_{2\Delta_\phi + 2\Delta_\psi - p}^{(p)}} \frac{\pi^{\frac{p}{2}}}{\Gamma\left(\frac{p}{2}\right)} \frac{1}{\varepsilon} + \text{O}(\varepsilon^0)\right) \frac{1}{|x|^{4\Delta_\psi}}
%\end{equation}
%and the same behavior is obtained when $y_1,y_2 \to x$. Hence we obtain
\begin{equation}
\begin{aligned}
	 \begin{tikzpicture}[scale=0.4, baseline=-0.1cm]
	 \draw[double,thick,blue] (-4,0)--(4,0);
	  \draw[thick, dotted, black]    (-3,0) to[out=90,in=180] (-1,1.5) to[out=0,in=90]  (1,0);
	  \draw[thick, dotted, black]    (-3,0) to[out=90,in=180] (-2,0.8) to[out=0,in=90]  (-1,0);
	   \draw[thick, black]    (-1,0) to[out=90,in=180] (0,0.8) to[out=0,in=90]  (1,0);
	   \draw[thick, dotted, black]    (1,0) to[out=270,in=180] (2,-0.6) to[out=0,in=270]  (3,0);
	  \draw[thick, dotted, black]    (-1,0) to[out=9270,in=180] (1,-1.2) to[out=0,in=270]  (3,0);
	  	 \draw[blue, fill=blue]   (-3,0) circle (4pt);
	  	  \draw[blue, fill=blue]   (3,0) circle (4pt);
	  	  \draw[blue, fill=white]   (-1,0) circle (4pt);
	  	  \draw[blue, fill=white]   (1,0) circle (4pt);
	  	  \node[below] at (-3,0) {$\hat{\psi}^2$};
	  	   \node[below] at (3.4,0) {$\hat{\psi}^2$};
	\end{tikzpicture} & = \frac{2 \mathcal{N}_\phi^{\,2} \! \left(\mathcal{N}_{\hat{\psi}}^{\,2}\right)^{4} \!  g_0^2}{|\tau|^{4\hat{\Delta}_{\hat{\psi}}}} \! \left(\frac{w_{2\Delta_\phi}^{(p)} w_{2\hat{\Delta}_{\hat{\psi}}}^{(p)}}{w_{2\Delta_\phi + 2\hat{\Delta}_{\hat{\psi}} - p}^{(p)}} \frac{2\pi^{\frac{p}{2}}}{\Gamma\!\left(\frac{p}{2}\right)} \frac{1}{\varepsilon} + \text{O}(\varepsilon^0)\right) \,.
 \end{aligned}
\end{equation}
This divergence must be canceled by introducing the wavefunction renormalization  $Z_{\hat{\psi}^2}$. Then, using \eqref{betanonlocal}, we can extract the beta function and the fixed point $g_{*}$
\begin{equation}
\begin{split}
Z_{\hat{\psi}^2} = 1 + g^2 \frac{\Gamma\!\left(\frac{p}{2} - \Delta_\phi\right)}{2^{d} \pi^{\frac{d}{2}} \Gamma\!\left(\frac{p}{2}\right) \Gamma\!\left(\frac{d}{2} - \Delta_\phi\right)\varepsilon} + \text{O}\left(g^4\right)\,,
\\ \beta_g = - \varepsilon g - \frac{\Gamma\!\left(\frac{p}{2} - \Delta_\phi\right)}{2^{d-1} \pi^{\frac{d}{2}} \Gamma\!\left(\frac{p}{2}\right) \Gamma\!\left(\frac{d}{2} - \Delta_\phi\right)} g^3 + \text{O}(g^4)  \,,
\\ g_{*}^2 = - \frac{2^{d-1} \pi^{\frac{d}{2}} \Gamma\!\left(\frac{p}{2}\right) \Gamma\!\left(\frac{d}{2} - \Delta_\phi\right)}{\Gamma\!\left(\frac{p}{2} - \Delta_\phi\right)} \, \varepsilon + \text{O}(\varepsilon^2)\,.
\end{split}
\end{equation}
Notice that the fixed-point coupling $g_{*}$ is not necessarily real. Interestingly, when $d=p+2$, we find that $g_{*}^2>0$ for any $\Delta_\phi$ above the unitarity bound. In the general case, only certain ranges of $\Delta_\phi$  yield a real fixed point.

Using \eqref{onepointphi2nonlocal}, we can also extract the one-point function of $\phi^2$ 
\begin{equation}
a_{\phi^2} = -\frac{9\,\Gamma \!\left(\frac{p}{2}\right)^2 \Gamma\!\left(\Delta_\phi-\frac{p}{2}\right) \Gamma\!\left(\frac{p-\Delta_\phi}{2}\right)^2}{16\,\Gamma\!\left(\frac{\Delta_\phi}{2}\right)^2 \Gamma\!\left(\frac{p}{2}-\Delta_\phi\right)(p-1)!}\, \varepsilon +\text{O}\left(\varepsilon^2\right)\,.
\end{equation}

In summary, we find that the interaction term $\phi \hat{\psi}^2$ provides a non-trivial defect CFT for some values of $p$, $d$, and $\Delta_\phi$. Specifically, for $d=3$ and $p=1$, we find a perturbative fixed point at
\begin{equation}
 g_{*}^2= \pi^2\varepsilon + \text{O}(\varepsilon^2)\,.
\end{equation}
Notice that we have not taken into account the bulk interaction here; therefore, we explicitly constructed an example of a non-trivial conformal defect in three-dimensional GFF theory. In this picture, $\Delta_\phi$ is a free parameter.

%In this picture one should interpret the perturbation as a modification in the dimension of the field $\phi$ (and the field $\psi$). For $\Delta_{\phi}=\frac{3}{4}$ the fixed point is trivial, but when we move away from that value we have a non-trivial interacting fixed point.

One may also wonder what the effect of introducing the bulk interaction would be. We can easily see that this effect would only contribute at higher orders in $\varepsilon$ for the beta function. However, in this case, $\Delta_\phi$ is no longer a free parameter, because we must take $\Delta_\phi=(d-\varepsilon)/4$.\,\footnote{Up to an arbitrary positive coefficient in front of $\varepsilon$, as it was discussed at the beginning of Section \ref{non-local}. }
For $d=3$, the fixed point with $p=1$ is the only real one, since $p=2$ yields a negative value for $g_*^2$.
 %Indeed to create a bulk vertex we need four scalar lines, which can only appear out of a defect interaction. Therefore the first diagram would appear at order $\lambda g^4$.

\subsection{$(a,b)=(2,1)$}

The second case we analyze is $(a,b)=(2,1)$, with interaction term $ \frac{g_0}{2}  \int d^p \tau \phi^2 \hat{\psi}$. Again, we first consider the free bulk case. The operator $\hat{\mathcal{O}}$ is given by $\hat{\mathcal{O}}= \phi \hat{\psi}$, and the tree-level contribution to its bare two-point function is given by
 \begin{equation}
 	\begin{tikzpicture}[scale=0.4, baseline]
	 \draw[double,thick,blue] (-3,0)--(3,0);
	  \draw[thick, dotted, black]    (-2,0) to[out=90,in=180] (0,2) to[out=0,in=90]  (2,0);
	   \draw[thick, black]    (-2,0) to[out=90,in=180] (0,1.3) to[out=0,in=90]  (2,0);
	  	 \draw[blue, fill=blue]   (-2,0) circle (4pt);
	  	  \draw[blue, fill=blue]   (2,0) circle (4pt);
	  	   \node[below] at (-2,0) {$\phi \hat{\psi}$};
	  	   \node[below] at (2,0) {$\phi \hat{\psi}$};
	\end{tikzpicture}
	 \, = \frac{\mathcal{N}_\phi^{\,2} \mathcal{N}_{\hat{\psi}}^{\,2}}{|\tau|^{2(\Delta_\phi + \hat{\Delta}_{\hat{\psi}})}}\,.
 \end{equation}
Again, at one loop there is only one divergent diagram , whose divergent can be computed as before
\begin{equation}
\begin{aligned}
	 \begin{tikzpicture}[scale=0.4, baseline]
	 \draw[double,thick,blue] (-4,0)--(4,0);
	  \draw[thick, dotted, black]    (-3,0) to[out=90,in=180] (-1,1.5) to[out=0,in=90]  (1,0);
	  \draw[thick, black]    (-3,0) to[out=90,in=180] (-2,0.8) to[out=0,in=90]  (-1,0);
	   \draw[thick, black]    (-1,0) to[out=90,in=180] (0,0.8) to[out=0,in=90]  (1,0);
	   \draw[thick, black]    (1,0) to[out=270,in=180] (2,-0.6) to[out=0,in=270]  (3,0);
	  \draw[thick, dotted, black]    (-1,0) to[out=9270,in=180] (1,-1.2) to[out=0,in=270]  (3,0);
	  	 \draw[blue, fill=blue]   (-3,0) circle (4pt);
	  	  \draw[blue, fill=blue]   (3,0) circle (4pt);
	  	  \draw[blue, fill=white]   (-1,0) circle (4pt);
	  	  \draw[blue, fill=white]   (1,0) circle (4pt);
	  	  \node[below] at (-3,0) {$\phi \hat{\psi}$};
	  	   \node[below] at (3.4,0) {$\phi \hat{\psi}$};
	\end{tikzpicture} & = \frac{\left(\mathcal{N}_\phi^{\,2}\right)^3 \!\left(\mathcal{N}_{\hat{\psi}}^{\,2}\right)^2}{{|\tau|^{2\Delta_\phi + 2\hat{\Delta}_{\hat{\psi}}}}} \left(\frac{w_{2\Delta_\phi}^{(p)} w_{2\hat{\Delta}_{\hat{\psi}}}^{(p)}}{w_{2\Delta_\phi + 2\hat{\Delta}_{\hat{\psi}} - p}^{(p)}} \frac{2\pi^{\frac{p}{2}}}{\Gamma\!\left(\frac{p}{2}\right)} \frac{1}{\varepsilon} + \text{O}(\varepsilon^0)\right)\,.
 \end{aligned}
\end{equation}

\noindent This yields the following renormalization factor,  beta functions and fixed point
\begin{equation}
\begin{split}
Z_{\phi \hat{\psi}} = 1 + \frac{\Gamma\!\left(\frac{p}{2} - \Delta_\phi\right)^2 g^2}{2^{2d-p-1} \pi^{d - \frac{p}{2}}\Gamma\!\left(\frac{p}{2}\right)\Gamma\!\left(\frac{d}{2} - \Delta_\phi\right)^2 \varepsilon} + \text{O}\left(g^4\right)\,,
\\ \beta_g = - \varepsilon g - \frac{\Gamma\!\left(\frac{p}{2} - \Delta_\phi\right)^2}{2^{2(d-1)-p} \pi^{d - \frac{p}{2}}\Gamma\!\left(\frac{p}{2}\right)\Gamma\!\left(\frac{d}{2} - \Delta_\phi\right)^2} g^3 + \text{O}\left(g^4\right)\,,
\\ g_{*}^2 = - \frac{2^{2(d-1)-p} \pi^{d - \frac{p}{2}}\Gamma\!\left(\frac{p}{2}\right)\Gamma\!\left(\frac{d}{2} - \Delta_\phi\right)^2}{\Gamma\!\left(\frac{p}{2} - \Delta_\phi\right)^2} \varepsilon + \text{O}(\varepsilon^2)\,.
\end{split}
\end{equation}
In this case, we find that the fixed-point value $g_*^2$ is always negative. Therefore, we cannot find any real fixed point.
Moreover, as in the previous case, the situation does not change at leading order when the bulk interactions are turned on.

%\noindent These allow us to derive the following CFT datum:

%\begin{equation}
%a_{\phi^2} = - \frac{9 \Gamma\left(\Delta_\phi\right) \Gamma\left(\Delta_\phi - \frac{p}{2}\right)\Gamma\left(\frac{p}{2}\right)^2 \Gamma\left(p-2\Delta_\phi\right)}{16\Gamma\left(2\Delta_\phi - \frac{p}{2}\right)\Gamma\left(\frac{p}{2} - \Delta_\phi\right)^2(p-1)!} \varepsilon + \mathcal{O}\left(\varepsilon^2\right)
%\end{equation}

\subsection{$(a,b) = (2,2)$}

The last case that we analyze explicitly is $(a,b) = (2,2)$. The interaction term is $ \frac{g_0}{2}  \int d^p \tau \phi^2 \hat{\psi}^2$. In this case the operator $\hat{\mathcal{O}}$ is $\hat{\mathcal{O}} = \phi \hat{\psi}^2$, and the tree-level bare two-point function is is given by the following diagram  \begin{equation}
 	 \begin{tikzpicture}[scale=0.4, baseline=-0.1cm]
	 \draw[double,thick,blue] (-3,0)--(3,0);
	  \draw[thick, dotted, black]    (-2,0) to[out=90,in=180] (0,2) to[out=0,in=90]  (2,0);
   \draw[thick, dotted, black]    (-2,0) to[out=90,in=180] (0,2.5) to[out=0,in=90]  (2,0);
	   \draw[thick, black]    (-2,0) to[out=90,in=180] (0,1.3) to[out=0,in=90]  (2,0);
	  	 \draw[blue, fill=blue]   (-2,0) circle (4pt);
	  	  \draw[blue, fill=blue]   (2,0) circle (4pt);
	  	   \node[below] at (-2,0) {$\phi \hat{\psi}^2$};
	  	   \node[below] at (2,0) {$\phi \hat{\psi}^2$};
	\end{tikzpicture}
	 \, = \frac{2 \mathcal{N}_\phi^{\,2} \!\left(\mathcal{N}_{\hat{\psi}}^{\,2}\right)^{2}}{|x|^{2\Delta_\phi + 4\hat{\Delta}_{\hat{\psi}}}}\,.
 \end{equation}
In this case we have a single one-loop contribution
\begin{equation}
\begin{aligned}
 	  \begin{tikzpicture}[scale=0.4, baseline=-0.1cm]
	 \draw[double,thick,blue] (-4,0)--(4,0);
	  \draw[thick, black]    (-3,0) to[out=90,in=180] (-1.5,1) to[out=0,in=90]  (0,0);
   \draw[thick, dotted, black]    (-3,0) to[out=90,in=180] (-1.5,1.8) to[out=0,in=90]  (0,0);
	   \draw[thick, dotted, black]    (-3,0) to[out=90,in=180] (0,3) to[out=0,in=90]  (3,0);
	  \draw[thick, black]    (0,0) to[out=90,in=180] (1.5,1) to[out=0,in=90]  (3,0);
   \draw[thick, dotted, black]    (0,0) to[out=90,in=180] (1.5,1.8) to[out=0,in=90]  (3,0);
	  	 \draw[blue, fill=blue]   (-3,0) circle (4pt);
	  	  \draw[blue, fill=blue]   (3,0) circle (4pt);
	  	  \draw[blue, fill=white]   (0,0) circle (4pt);
	  	  \node[below] at (-3,0) {$\phi \hat{\psi}^2$};
	  	   \node[below] at (3,0) {$\phi \hat{\psi}^2$};
	\end{tikzpicture}
 \, &= -\frac{4 g_0\!\left(\mathcal{N}_\phi^{\,2}\right)^2 \!\left(\mathcal{N}_{\hat{\psi}}^{\,2}\right)^3}{{|\tau|^{4\Delta_\phi + 6 \hat{\Delta}_{\hat{\psi}} - p}}}  \frac{\left(w_{2\Delta_\phi + 2\hat{\Delta}_{\hat{\psi}}}^{(p)}\right)^2}{w_{4\Delta_\phi + 4 \hat{\Delta}_{\hat{\psi}} - p}^{(p)}} \,.
\end{aligned}
\end{equation}
This diagram diverges, and its pole can be canceled by introducing the wavefunction renormalization $Z_{\phi \hat{\psi}^2}$. We find
\begin{equation}
\begin{split}
Z_{\phi \hat{\psi}^2} = 1 - \frac{\Gamma\!\left(\frac{p}{2} - \Delta_\phi\right) g}{2^{d-2} \pi^{\frac{d}{2}}\Gamma\!\left(\frac{p}{2}\right) \Gamma\!\left(\frac{d}{2} - \Delta_\phi\right) \varepsilon} + \text{O}\left(g^2\right)\,,
\\ \beta_g = - \varepsilon g + \frac{\Gamma\!\left(\frac{p}{2} - \Delta_\phi\right)}{2^{d-2} \pi^{\frac{d}{2}}\Gamma\!\left(\frac{p}{2}\right) \Gamma\!\left(\frac{d}{2} - \Delta_\phi\right)} g^2 + \text{O}\left(g^3\right)\,,
\\ g_{*} = \frac{2^{d-2} \pi^{\frac{d}{2}}\Gamma\!\left(\frac{p}{2}\right) \Gamma\!\left(\frac{d}{2} - \Delta_\phi\right)}{\Gamma\!\left(\frac{p}{2} - \Delta_\phi\right)} \varepsilon + \text{O}\left(\varepsilon^2\right)\,.
\end{split}
\end{equation}
In this case, we have a fixed-point value for the coupling $g_*$ (and not for $g_*^2$), so we do not need to worry about its reality. Still, as we know from Section \ref{subsubnonloc}, the value $a=2$ is only allowed for $p\geq 2$. Therefore, the physically interesting case is $p=2$, $d=3$, which gives
\begin{equation}
 g_*=\sqrt{2\pi}\ \Gamma \! \left(\frac{3}{4}\right)^2 \varepsilon\,.
\end{equation}
Using equation \eqref{onepointphi2nonlocal} we can easily extract the one-point function of $\phi^2$
\begin{equation}
a_{\phi^2} = \frac{\Gamma\!\left(\Delta_\phi - \frac{p}{2}\right) \Gamma\!\left(\frac{p}{2}\right)^3}{2\,\Gamma\!\left(\Delta_\phi\right)(p-1)!} \varepsilon^2 + \text{O}\left(\varepsilon^3\right)\,.
\end{equation}

As in the discussion of the $(a,b)=(1,2)$ case, we have not turned on the bulk interaction yet, meaning that this example provides a non-trivial surface defect in three-dimensional GFF theory.
If we were to include the bulk interaction, it would still not affect the leading-order computation. %Indeed, the first diagram with a bulk interaction contributing to the two-point function of $\phi \psi^2$ would require at least one defect vertex to produce a total of four scalar line joining in the bulk. Therefore, it would give a contribution of order $\lambda g$ which would contribute to the fixed-point value $g_*$ at order $\varepsilon^2$.

\subsection{Summary}

We can summarize our findings for the physically interesting case of $d = 3$. In this case, we can have two types of defects: lines and surfaces.
\begin{itemize}
 \item For $p = 1$, the allowed interaction terms are of the form $\phi \hat{\psi}^b$. We analyzed in detail the case $b = 2$ in GFF theory, finding the existence of a non-trivial line defect. Moreover, we argued that, at leading order, the bulk interaction does not affect this result.
 \item For $p = 2$, we can have interactions $\phi \hat{\psi}^b$ and $\phi^2 \hat{\psi}^b$. We analyzed in detail the two cases with $b = 2$, finding a non-trivial defect only for the second interaction.
Again, at leading order, the bulk interaction does not affect this result.
\end{itemize}
Of course, for higher values of $b$, other potentially interesting defect examples could arise. It would also be interesting to investigate whether there are additional constraints on the allowed values of $b$. We leave this analysis for future work.

 \section{Defects in the long-range $O(N)$ model close to four dimensions} \label{epsexp}

In this section, we discuss the generalization of two simple defects that can be constructed in the short-range $O(N)$ model, starting from the UV theory in $d=4$ and expanding around $d = 4 - \hat{\varepsilon}$. Unlike the previous section, where we worked at fixed $d$ by varying the dimension of the scalar field $\Delta_{\phi}$, here we need to expand around $d=4$ for the defect action to be weakly relevant.

To explore the LRI fixed points, in Section \ref{LRImodel} we introduced a parameter $0<\kappa<1$, with $\sigma = 2 - \frac{(1 - \kappa)\hat{\varepsilon}}{2}$. 
This parameterizes a straight line in the $(d, \sigma)$ plane, where $\kappa$ selects a particular direction in the phase space shown in Figure \ref{phases}. 
In general, the LRI fixed points in the $(d, \sigma)$ plane are characterized by the exact relation $\Delta_{\phi} = \frac{d - \sigma}{2}$. When performing calculations near the local theory in $d=4$, non-vanishing contributions to the wavefunction renormalization and the anomalous dimension of the field $\phi$ arise. Therefore, by enforcing the condition above, we obtain the following expression for $\sigma$:
\begin{equation}
\sigma = d - 2\Delta_{\phi} = 2 + \frac{\kappa - 1}{2} \hat{\varepsilon} + 2\gamma_{\phi}(\kappa, \hat{\varepsilon}),
\end{equation}
where $\gamma_{\phi}(\kappa, \hat{\varepsilon})$ is the anomalous dimension of the field $\phi$, which starts at $\text{O}(\hat{\varepsilon}^2)$. This parametrizes a one-parameter family of trajectories, ranging from $\kappa = 0$, where $\gamma_{\phi}(0, \hat{\varepsilon}) = 0$ and $\sigma = d/2$, to $\kappa = 1$, where $\gamma_{\phi}(1, \hat{\varepsilon}) = \gamma_{\text{SRI}}$ and the trajectories approach the upper bound of the LRI region, as shown in Figure \ref{phases}.

\begin{figure}[htbp]
    \centering
    \includegraphics[width=0.8\linewidth] {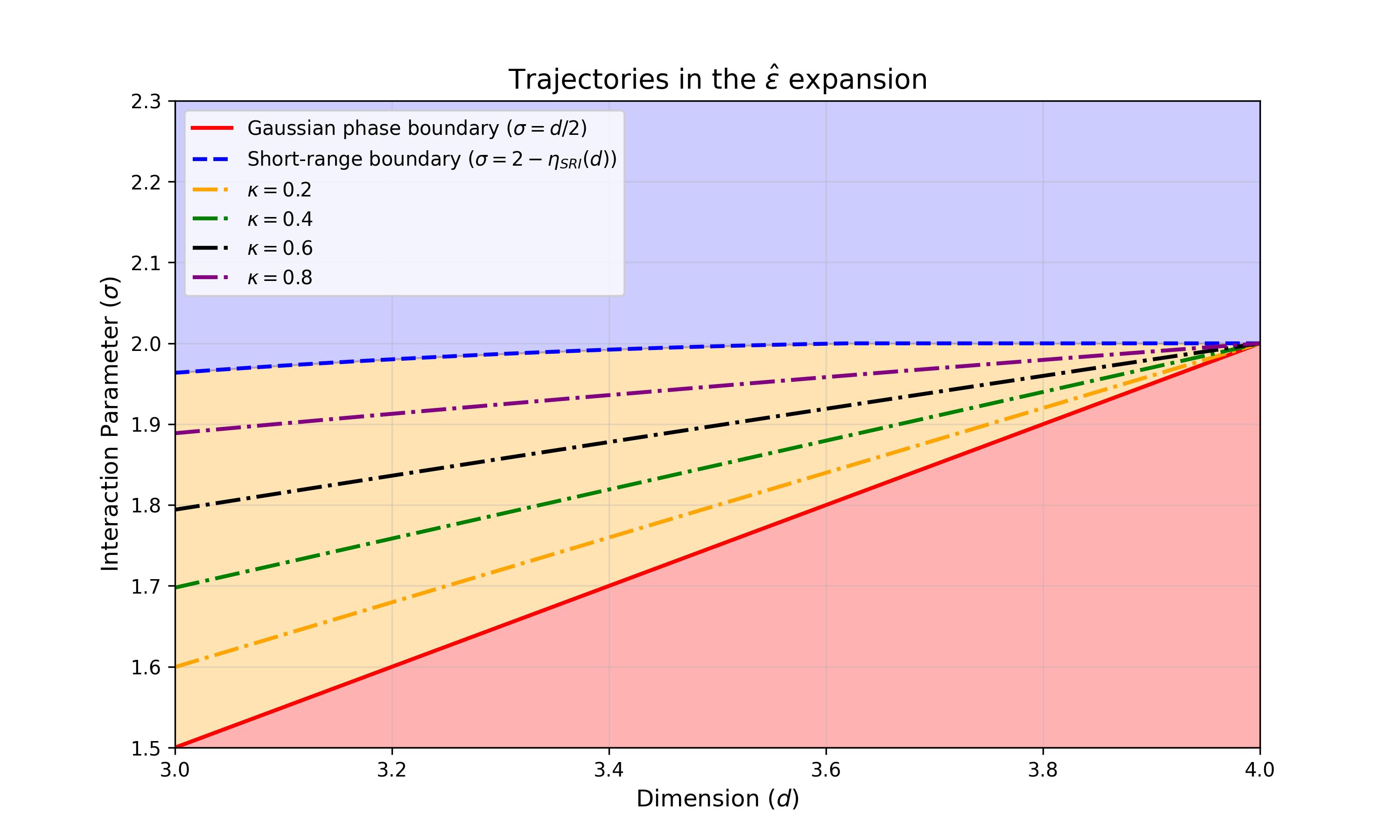}
    \caption{The trajectories $\sigma(d)$ such that the condition $\sigma=d-2\Delta_{\phi}$ is enforced.}
    \label{comparison}
\end{figure}

By making a simple generalization of the computation for the SRI (or $\kappa = 1$), we can obtain these trajectories at order $\varepsilon^2$. Note that, at this stage, we are only considering the homogeneous theory, without any defect.
The first contribution to the renormalization of the field $\phi$ appears at two loops, and it gives:
\begin{equation}
Z_\phi=1-\frac{\left(N+2\right)\lambda^2}{18\left(4\pi\right)^4\left(1+3\kappa\right)\hat{\varepsilon}} + \text{O}(\lambda^3)\,,
\end{equation}
while the coupling renormalization yields
\begin{equation}
Z_\lambda = 1 + \frac{N+8}{3 \kappa \,\hat{\varepsilon}} \frac{\lambda}{(4\pi)^2} + \left(\frac{(N+8)^2}{9 \kappa^2 \hat{\varepsilon}^2} - \frac{5N + 22 + \kappa(13N+62)}{9(1+3\kappa)\kappa \,\hat{\varepsilon}}\right) \frac{\lambda^2}{(4\pi)^4} + \text{O}(\lambda^3)\,.
\end{equation}
This leads to the following beta function:
\begin{equation}
\beta_\lambda = - \kappa \,\hat{\varepsilon} \,\lambda + \frac{N+8}{3} \frac{\lambda^2}{(4\pi)^2} - \frac{10N+44+\kappa(26N+124)}{9(1+3\kappa)} \frac{\lambda^3}{(4\pi)^4} + \text{O}(\lambda^4)\,,
\end{equation}
which admits a non-trivial (Wilson-Fisher) fixed point at
\begin{equation}\label{bulkfixedpoint}
\frac{\lambda_\ast}{\left(4\pi\right)^2}=\frac{3}{N+8}\kappa\,\hat{\varepsilon}+\frac{6}{\left(N+8\right)^3}\frac{5N+22+\kappa\left(13N+62\right)}{1+3\kappa}\kappa^2\hat{\varepsilon}^2+\text{O}\left(\hat{\varepsilon}^3\right)\,.
\end{equation}
From these computations one can immediately derive the anomalous dimension $\gamma_\phi$ of $\phi_a$ at the IR fixed point
\begin{equation}
\gamma_\phi = \beta_\lambda \frac{\partial \ln Z_\phi}{\partial \lambda} \biggr \rvert_{\lambda=\lambda_{*}} = \frac{\kappa^3(N+2)}{(1+3\kappa)(N+8)^2} \hat{\varepsilon}^2 +\text{O}\left(\hat{\varepsilon}^3\right)\,.
\end{equation}
This leads to the following family of curves in the $(d,\sigma)$ plane
\begin{equation}
 \sigma= 2+\frac{\kappa-1}{2} \hat{\varepsilon} -\frac{2  \kappa ^3 (N+2)}{(3 \kappa +1) (N+8)^2} \hat{\varepsilon} ^2+\text{O}\left( \hat{\varepsilon} ^3\right)\,.
\end{equation}
Some of these trajectories are shown in Figure \ref{comparison}.
As a consistency check, one can easily verify that at $\kappa = 1$ one
retrieves the expressions for the short-range $O(N)$ model
\begin{equation}
\sigma = 2-\frac{\left(N+2\right)}{2\left(N+8\right)^2}\hat{\varepsilon}^2=2-\eta_{\text{SRI}}\,.
\end{equation}

\noindent Another useful check, which may be more trivial in flows that do not depend on an extra parameter like $\kappa$, is the stability of the bulk IR fixed point. This involves studying the sign of the derivative of $\beta_\lambda$ at $\lambda_{*}$ and ensuring it is positive
\begin{equation}
\beta_\lambda'(\lambda_{*}) = \kappa \, \hat{\varepsilon} - \frac{2\left(5N+22+\kappa(13N+62)\right)}{(1+3\kappa)(N+8)^2} \kappa^2 \hat{\varepsilon}^2 + \text{O}(\hat{\varepsilon}^3)\,.
\end{equation}
This leads to the following consistency condition:
\begin{equation}
\hat{\varepsilon} < \hat{\varepsilon}_{\text{thresh}} = \frac{(1+3\kappa)(N+8)^2}{2 \kappa \left(5N+22+\kappa(13N+62)\right)}\,.
\end{equation}
For $0<\kappa <1$, this threshold value is always greater than 1, meaning that this stability condition is never a concern for our perturbative treatment.

\subsection{Localized magnetic field}\label{localizedmagneticfield}

We consider now the insertion of defects in the setup described above. The simplest defect one can consider is the localized magnetic field
\begin{equation}
 S=S_0+h_0 \int d\tau \phi_1 (\tau)\,,
\end{equation}
where a single field $\phi_1$ is integrated along a line. We would like to analyze the renormalization of the coupling $h$ in the $\hat{\varepsilon}$-expansion. We follow the method described in \cite{Cuomo:2021rkm} imposing finiteness of the one-point function of $\phi_1$. The computation is analogous to the local case, up to the appearence of the new parameter $\kappa$. The relevant diagrams are shown in Appendix \ref{locmagdiag}. The final result for the coupling renormalization $Z_h$ is
\begin{equation}\label{Zhlocmag}
\begin{aligned}
Z_h = 1 + \tfrac{h^2}{3(1+3\kappa) \hat{\varepsilon}} \tfrac{\lambda}{(4\pi)^2} + & \left[\tfrac{(N+2)h}{18(1+3\kappa)\hat{\varepsilon}}+ \left(\tfrac{1}{\hat{\varepsilon}^2} - \tfrac{1 + 3\kappa^2 - (1-\kappa)(\gamma_E - \log 4\pi)}{4\kappa \hat{\varepsilon}}\right)\tfrac{2(N+8)h^2}{9(1+3\kappa)(1+5\kappa)} \right.
\\ & \left. + \left(\tfrac{2}{(1+3\kappa)\hat{\varepsilon}^2} - \tfrac{1}{\hat{\varepsilon}}\right)\tfrac{h^4}{12(1+3\kappa)}\right]\tfrac{\lambda^2}{(4\pi)^4} + \text{O}(\lambda^3)\,.
\end{aligned}
\end{equation}
This subsequently yields the following beta function\,\footnote{
One should not be worried about the appearence of terms like $\gamma_E$ or $\log 4 \pi$, since two-loop coefficients of the beta function for multiple couplings are scheme dependent.}
\begin{equation}\label{betahlamlocmag}
\begin{aligned}
\beta_h &=-\frac{\hat{\varepsilon}\left(1+\kappa\right)}{4}h+\frac{\lambda}{\left(4\pi\right)^2}\frac{h^3}{6}+\frac{\lambda^2}{\left(4\pi\right)^4}\bigg(\frac{\left(N+2\right)\kappa h}{9\left(1+3\kappa\right)} \\
 & \quad \quad-\frac{\left(N+8\right)\left(1+3\kappa^2-\left(1-\kappa\right)\left(\gamma_E-\log{\left(4\pi\right)}\right)\right)}{36\kappa\left(1+3\kappa\right)}h^3 -\frac{h^5}{12}\bigg)
+ \text{O}(\lambda^3)\,.
\end{aligned}
\end{equation}
This beta function reduces to the short range result in the limit $\kappa \to 1$  (see for instance \cite{Cuomo:2021kfm}). The opposite limit, $\kappa\to 0$, is subtle because in \eqref{Zhlocmag} the single pole in $\hat{\varepsilon}$ is accompanied by a single pole in $\kappa=0$, leading to a singular limit $\kappa\to 0$ for the beta function \eqref{betahlamlocmag}. This pole is cancelled when we set the bulk coupling to the fixed point $\lambda_*$ in \eqref{bulkfixedpoint}.
%\begin{equation}
%\begin{aligned}
%& \beta_h(\lambda_{*}) = -\hat{\varepsilon} \left( \frac{(1+\kappa)h}{4} - \frac{\kappa h^3}{2(N+8)} \right)+\hat{\varepsilon}^2 \left(\frac{(N+2)\kappa^3h}{(1+3\kappa)(N+8)^2}\right.
%\\ & + \frac{4\kappa(5N+22+\kappa(13N+62)) - (N+8)^2 \left(1 + 3\kappa^2 + (1-\kappa)(\log (4 \pi) - \gamma_E)\right)}{4(1+3\kappa)(N+8)^3}\kappa h^3
%\\ &\left. - \frac{3\kappa^2 h^5}{4(N+8)^2}\right)
%+ \text{O}(\hat{\varepsilon}^3)
%\end{aligned}
%\end{equation}
In that case, the limit $\kappa \to 0$ trivially leads to $\beta_h=-\hat{\varepsilon}h/4$, consistently with a Gaussian fixed point. For general $\kappa$, instead one has a non-trivial defect fixed point
\begin{equation}
\begin{aligned}
h_{*}^2 &= \frac{(N+8)(1+\kappa)}{2\kappa} 
+ \frac{\left(5 + 2(1-\kappa^2) (\log(4\pi)-\gamma_E)\right)(N+8)}{8\kappa(1+3\kappa)} \hat{\varepsilon}
\\ &+ \frac{(15\kappa^2 + 27\kappa + 17)  N^2 + 8(15\kappa^2 + 36\kappa+29) N+48 (9\kappa^2 + 22\kappa+19)}{8(1+3\kappa)(N+8)} \hat{\varepsilon} + \text{O}(\hat{\varepsilon}^2)\,.
\end{aligned}
\end{equation}
We now proceed to compute some defect observables at this fixed point.
%As before, one can also study the stability of the fixed point. Requiring a positive derivative of the $\beta$-function at the fixed point yields the following numerical requirement:
%\begin{equation}
%\frac{1+\kappa}{2} \geq \left(\frac{(9N^2+160N+608)\kappa^3}{8(1+3\kappa)(N+8)^2} - \frac{3}{8} \frac{7\kappa^2+5\kappa+1+\left(1-\kappa^2\right)\left(\ln{4\pi}-\gamma_E\right)}{1+3\kappa}\right) \hat{\varepsilon}
%\end{equation}
%Incidentally, this implies that the scaling dimension of $\phi$ on the defect is positive (see next section), hence satisfying a unitarity bound.

\paragraph{Scaling dimension.} The first observable that we can compute is the scaling dimension of the defect field $\hat{\phi}_1$. This is just a derivative of the beta function: $\Delta_{\hat{\phi}}=1+ \frac{\partial \beta_h}{\partial h}|_{h=h_*}$, and it is
\begin{equation}
\Delta_{\hat{\phi}} = 1 + \frac{1+\kappa}{2} \hat{\varepsilon} - \frac{1}{8}\left( 3 + \kappa \left( 	 3(\kappa+2)+\frac{16 \kappa^2 (N+2)}{(1+3\kappa)(N+8)^2}\right)\right) \hat{\varepsilon}^2+\text{O}(\hat{\varepsilon}^3)\,.
\end{equation}
It is important to notice that the limit $\kappa\to 0$ should not be considered, since there is no defect fixed point in that direction. The limit $\kappa \to 1$, instead, perfectly reproduces the existing results \cite{Cuomo:2021kfm}.

\paragraph{One-point function.} The next piece of defect CFT data we consider is the one-point function of the bulk field $\phi_1$, which is the observable we used to renormalize the defect coupling. Defining
\begin{equation}\label{onepointclosetofour}
\langle \phi_a(x) \rangle = \delta_{a,1} \frac{\mathcal{N}_\phi a_{\phi}}{|x_{\perp}|^{\Delta_\phi}}\,,
\end{equation}
and inserting the renormalized coupling in the diagrams in Appendix \ref{locmagdiag}, we find
\begin{equation}\label{onepointcoefficientclosetofour}
\begin{aligned}
a_{\phi}^2 &= \frac{(\kappa+1) (N+8)}{8 \kappa} + \bigg(\frac{-15 \kappa ^2-18 \kappa -5}{4 (3 \kappa +1)}+\frac{3 \left(9 \kappa ^2+10 \kappa +3\right)}{2 (3 \kappa +1) (N+8)} \\
& \quad \quad \quad  \quad \quad +\frac{(3 \kappa -1) (\kappa +1) (N+8)}{32 \kappa }
 +\frac{(\kappa +1)^2 (N+8) \log (2)}{16 \kappa }\bigg) \hat{\varepsilon}+\text{O}\left(\hat{\varepsilon}^2\right)\,.
\end{aligned}
\end{equation}
When $\kappa \to 1$, we recover the result of \cite{Cuomo:2021kfm}.

\paragraph{$g$-function.} Another important defect observable is the $g$-function of this DCFT, which obeys a monotonicity theorem under RG flow \cite{Cuomo:2021rkm}. The computation is analogous to the one in \cite{Cuomo:2021kfm}, with the important difference that we keep a generic value of $\Delta_{\phi}$ leading to a dependence on $\kappa$. For a free bulk theory, we can compute the defect expectation value of the circular defect exactly either by finding the classical solution to the equation of motion and computing the classical action or by resumming diagrams. The final result is
\begin{equation}
\begin{aligned}
\log g_{\text{free}} = \log \frac{Z_{\text{defect}}}{Z_0} = \frac{2^{1-d} \pi ^{\frac{3-d}{2}} \Gamma
   \left(\frac{1}{2}-\Delta _{\phi }\right) \Gamma
   \left(\Delta _{\phi }\right) R^{2-2 \Delta _{\phi
   }}}{\Gamma \left(1-\Delta _{\phi }\right) \Gamma
   \left(\frac{d}{2}-\Delta _{\phi }\right)}\,,
\label{free g}
\end{aligned}
\end{equation}
where $R$ is the radius of the circle. 
Adding the bulk interaction, we have two diagrams up to order $\hat{\varepsilon}$, which were solved in \cite{Cuomo:2021kfm} and we review in Appendix \ref{g-fun}. The final result is
\begin{equation}
\log g = - \frac{(1+\kappa)\hat{\varepsilon}}{16}h^2+\frac{\kappa h^4 \lambda_{*} }{192 \pi ^2 (3 \kappa+1)}+\text{O}\left(\lambda_{*}^2\right)\,,
\end{equation}
and using the value of the defect fixed point coupling $h_{*}$
\begin{equation}
\log g_{\text{IR}} = -\frac{(\kappa+1)^3 (N+8)}{32 \kappa (3 \kappa+1)} \hat{\varepsilon} +\text{O}\left(\hat{\varepsilon}^2\right) < 0 = \log g_{\text{UV}}\,,
\end{equation}
as one expects from the $g$-theorem.\,\footnote{More precisely, the quantity that is monotonic under RG flow is the defect entropy, defined by $s=(1-R\frac{\partial}{\partial R})\log g$. However, as explained in \cite{Cuomo:2021kfm}, $s$ and $\log g$ agree at leading order.}

\subsection{Surface defect}

We now turn to another class of defects, which was analyzed in the short-range model in \cite{Trepanier:2023tvb, Raviv-Moshe:2023yvq, Giombi:2023dqs}. The surface defect is realized by integrating $\phi^2$ over a two-dimensional plane\,\footnote{There is also a symmetry-breaking version of this surface defect obtained by the interaction $h^{ab}\phi_a \phi_b$, with a generic tensor coupling $h_{ab}$. However, for simplicity we only consider the $O(N)$ symmetric case.}
\begin{equation}
S =S_0 + h_0 \int d^2 \tau \, \phi_a \phi_a(\tau)\,,
\end{equation}
and, contrary to the localized magnetic field, it preserves the full $O(N)$ symmetry.

To renormalize the defect coupling we consider the one-point function of $\phi_a \phi^a$ (we need a $O(N)$ singlet to get a non-vanishing one-point function). When the bulk is free, the $\beta$ function can be computed exactly by going to momentum space and following the procedure described in \cite{Giombi:2023dqs} for resumming the diagrams
  \begin{equation}
  \begin{tikzpicture}[scale=0.4, baseline=-0.1cm]
	 \draw[double,thick,blue] (-3,0)--(3,0);
	  	  \draw[blue, fill=blue]   (0,0) circle (4pt);
          \draw[blue, fill=black]   (0,5) circle (4pt);
      \draw[thick, black]    (0,5) to[out=-45,in=90] (1,2.5) to[out=270,in=45]  (0,0);
      \draw[thick, black]    (0,5) to[out=225,in=90] (-1,2.5) to[out=270,in=135]  (0,0);
	  \node[above] at (0.2,5) {$\phi^2$};
	\end{tikzpicture}
 \hspace{0.4 cm} + \hspace{0.4 cm}
      \begin{tikzpicture}[scale=0.4, baseline=-0.1cm]
	 \draw[double,thick,blue] (-3,0)--(3,0);
     \draw[thick,black] (0,5)--(-1.7,0);
     \draw[thick,black] (0,5)--(1.7,0);
     \draw[thick,black] (-1.7,0) to[out=90,in=180] (0,1) to[out=0,in=90] (1.7,0);
	  	  \draw[blue, fill=blue]   (-1.7,0) circle (4pt);
           \draw[blue, fill=blue]   (1.7,0) circle (4pt);
          \draw[blue, fill=black]   (0,5) circle (4pt);
	  \node[above] at (0.2,5) {$\phi^2$};
	\end{tikzpicture}
\hspace{0.4 cm}+ \hspace{0.4 cm}
      \begin{tikzpicture}[scale=0.4, baseline=-0.1cm]
	 \draw[double,thick,blue] (-3,0)--(3,0);
     \draw[thick,black] (0,5)--(-1.7,0);
     \draw[thick,black] (0,5)--(1.7,0);
          \draw[thick,black] (-1.7,0) to[out=90,in=180] (-0.85,0.6) to[out=0,in=90] (0,0);
          \draw[thick,black] (0,0) to[out=90,in=180] (0.85,0.6) to[out=0,in=90] (1.7,0);
	  	  \draw[blue, fill=blue]   (-1.7,0) circle (4pt);
           \draw[blue, fill=blue]   (1.7,0) circle (4pt);
            \draw[blue, fill=blue]   (0,0) circle (4pt);
          \draw[blue, fill=black]   (0,5) circle (4pt);
	  \node[above] at (0.2,5) {$\phi^2$};
	\end{tikzpicture}
\hspace{0.4 cm}	+ \ \ \, \ldots  % = \sum_{n=0}^{+\infty} A_n
  \end{equation}
%\begin{equation}
%\begin{cases}
%A_0=-2h_0\frac{N}{\left(k^2+m^2\right)^\frac{\sigma}{2}\left(k^2+n^2\right)^\frac{\sigma}{2}}
%\\ A_{n+1} = t A_n
%\end{cases}
%\end{equation}

%\noindent with
%
%\begin{equation}
%t\left(k\right)=-2h_0\int{\frac{d^{d-2}p}{\left(2\pi\right)^{d-2}}\frac{1}{\left(k^2+p^2\right)^\frac{\sigma}{2}}}=-\frac{2h_0k^{d-2-\sigma}\Gamma\left(1-\frac{d-\sigma}{2}\right)}{\left(4\pi\right)^\frac{d-2}{2}\Gamma\left(\sigma\right)} = - \frac{2h}{(1+\kappa)\pi\hat{\varepsilon}} + \mathcal{O}(1)
%\end{equation}
This leads to the exact expression for the bare coupling
\begin{equation}
h_0 = \mu^{\frac{1+\kappa}{2}\hat{\varepsilon}}h \left(1 + \frac{2h}{(1+\kappa)\pi\hat{\varepsilon}} + \left(\frac{2h}{(1+\kappa)\pi\hat{\varepsilon}}\right)^2 + \dots \right) =\frac{\mu^{\frac{1+\kappa}{2}\hat{\varepsilon}}h}{1-\frac{2h}{\left(1+\kappa\right)\pi\hat{\varepsilon}}}\,,
\end{equation}
and consequently to the exact beta function
\begin{equation}\label{betasurffree}
\beta_h=-\frac{1+\kappa}{2}\hat{\varepsilon} h+\frac{h^2}{\pi}\,.
\end{equation}
This admits a non-trivial fixed point in $h_{*} = \frac{\pi(1+\kappa)}{2} \hat{\varepsilon}$.

Reintroducing the bulk coupling, we need to take into account the wavefunction renormalization of $\phi^2$, which will depend on the parameter $\kappa$. More generally, we can compute the renormalization of the field $\phi^n$ for any positive intrger $n$. This is given by the following bulk diagrams
\begin{equation}
\begin{aligned}
\begin{tikzpicture}[scale=0.4,baseline]

   \node[above] at (3.5,-0.3) {$\phi^n$};
   \node[above] at (-3.5,-0.3) {$\phi^n$};
  \draw[blue,fill=black] (-3,0) circle (2pt);
  \draw[blue,fill=black] (3,0) circle (2pt);
  % Draw the outer ellipse
  \draw[thick, black] (0,0) ellipse (3 and 1.5);
  
  % Draw the inner curves (propagators)
  \draw[thick, black] (-3,0) to[out=30, in=150] (3,0); % Upper curve
  \draw[thick, black] (-3,0) to[out=-30, in=-150] (3,0); % Lower curve
  
  % Central dots
  \node at (0, 0.4) {\large $\cdot$};
  \node at (0, 0) {\large $\cdot$};
  \node at (0, -0.4) {\large $\cdot$};
\end{tikzpicture}= \frac{Nn!N_\phi^n}{|x|^{2n\Delta_\phi}}\,, \hspace{4 cm}
\end{aligned}
\end{equation}

\begin{equation}
\begin{aligned}
& \begin{tikzpicture}[scale=0.3,baseline]
\node[above] at (6.3,-0.4) {$\phi^n$};
   \node[above] at (-6.3,-0.4) {$\phi^n$};
  % Outer ellipse (big, with semi-major axis 5.5 and semi-minor axis 2.3)
  \draw[thick, black] (0,0) ellipse (5.5 and 2.3);

  % Two inner ellipses (left and right, smaller, touching the outer ellipse)
  \draw[thick, black] (-2.75, 0) ellipse (2.75 and 1.0); % Left inner ellipse
  \draw[thick, black] (2.75, 0) ellipse (2.75 and 1.0);  % Right inner ellipse

    \draw[blue,fill=black] (-5.5,0) circle (4pt);
    \draw[blue,fill=black] (5.5,0) circle (4pt);
    \draw[blue,fill=black] (0,0) circle (4pt);
  
  % Single ellipse representing the top and bottom curves
  \draw[thick, black] (0, 0) ellipse (5.5 and 2.0);  % The full ellipse for the upper and lower curves

   % Central dots
  \node at (0, 0.7) {\large $\cdot$};
  \node at (0, 1.1) {\large $\cdot$};
  \node at (0, 1.5) {\large $\cdot$};

  % Central dots
  \node at (0, -0.7) {\large $\cdot$};
  \node at (0, -1.1) {\large $\cdot$};
  \node at (0, -1.5) {\large $\cdot$};
  
\end{tikzpicture}
=
\\ &=- \frac{N(N+2)n!n(n-1)N_\phi^{n+2}\lambda_0}{12} \frac{\left(w_{4\Delta_\phi}^{(d)}\right)^2}{w_{8\Delta_\phi - d}^{(d)}} \frac{1}{|x|^{(8+2n-4)\Delta_\phi - d}}\,.
\end{aligned}
\end{equation}
The second diagram diverges, and we can introduce the wavefunction renormalization\,%\footnote{While we don't need it here, we can easily derive the anomalous dimension of $\phi^n$: $\gamma_n(\lambda_{*}) = \beta_\lambda \frac{\partial \log Z_{\phi^n}}{\partial \lambda} =\frac{3n(n-1)(N+2)}{6(N+8)} \kappa \,\hat{\varepsilon} + \mathcal{O}\left(\hat{\varepsilon}^2\right)$.}
\begin{equation}
Z_{\phi^n} = 1 - \frac{(N+2)n(n-1)}{6\kappa\, \hat{\varepsilon}} \frac{\lambda}{(4\pi)^2} + \text{O}(\lambda^2)\,.
\end{equation}
%Evaluating the Feynman diagrams in the renormalized theory gives us the following normalization factor for the bulk two-point function of $\phi^n$:
%\begin{equation}
%N_{\phi^n} = \frac{Nn!}{16 \pi^4}\left[1 + \left(\gamma \kappa + (1-\kappa) \log 2 + \log \pi - \frac{n(n-1)}{2(N+8)} (N+2) (1 + \gamma(2\kappa-1) + 2(1-\kappa)\log 2 + \log \pi)\right) \hat{\varepsilon} + \mathcal{O}\left(\hat{\varepsilon}^2\right)\right]
%\end{equation}
The diagrams for the renormalization of the defect coupling $h$ are given in Appendix \ref{locmagdiag}. After taking into account the wavefunction renormalization $Z_{\phi^2}$, we find the following
\begin{equation}
\begin{split}
h_0 &= \mu^{\frac{1+\kappa}{2}\hat{\varepsilon}}h\left(\frac{1}{1-\frac{2h}{\left(1+\kappa\right)\pi\hat{\varepsilon}}} + \frac{(N+2)\lambda}{48 \pi^2 \kappa \,\hat{\varepsilon}}\right) + \text{O}\left(h^2\lambda,h\lambda^2,\lambda^3\right)\,,
\\ \beta_h &= -\frac{1+\kappa}{2}\hat{\varepsilon} h+\frac{h^2}{\pi}+\frac{N+2}{48\pi^2}h\lambda\,.
\end{split}
\end{equation}
At $\lambda=\lambda_*$, we find a non-trivial defect fixed point
\begin{equation}
h_\ast=\frac{N\left(1-\kappa\right)+8+4\kappa}{2\left(N+8\right)}\pi\hat{\varepsilon}+\text{O}\left(\hat{\varepsilon}^2\right)\,.
\end{equation}
As expected, for $\kappa = 1$ we recover the result of the short-range model \cite{Trepanier:2023tvb, Giombi:2023dqs, Raviv-Moshe:2023yvq}. In the opposite limit, and in contrast to the case of the localized magnetic field, there is no divergence at $\kappa = 0$. Instead, the fixed point gives $h_* = \frac{\pi}{2} \hat{\varepsilon}$, which is in agreement with the exact result found below for the Gaussian theory \eqref{betasurffree}. Another notable difference from the localized magnetic field is that the fixed point is perturbativly small in $\hat{\varepsilon}$, rather than being of order $\varepsilon^0$. 

Also in this case we can compute some interesting defect observables.

\paragraph{Scaling dimension.} As before, a derivative of the beta function for $h$ gives us access to the following defect scaling dimension at the fixed point:
\begin{equation}
\Delta_{\hat{\phi}^2} = 2 + \frac{8+4\kappa+N(1-\kappa)}{2(N+8)}\hat{\varepsilon} + \text{O}(\hat{\varepsilon}^2)\,,
\end{equation}
Once more, both limits $\kappa=1$ and $\kappa=0$ are well defined and they return the expected results for the short range and the Gaussian theory.

\paragraph{One point data.} The Feynman diagrams we used to renormalize the coupling also give us access to the one-point function $\langle \phi^2 \rangle$
\begin{equation}
a_{\phi^2} = - \frac{\sqrt{N} (4\kappa+N(1-\kappa)+8)}{4 \sqrt{2} (N+8)} \hat{\varepsilon} + \text{O}(\hat{\varepsilon}^2)\,,
\end{equation}

\paragraph{Defect free energy.} 
The last observable we discuss is the defect free energy $\mathcal{F}$, which, in the case of surface defects, is a divergent quantity that contains a universal logarithmic term related to the Weyl anomaly coefficient. Specifically, for a two-dimensional defect, the Weyl anomaly takes the form
\begin{equation}
\label{eq:defecttrace}
T^{\mu}_{~\mu}\Big|_{\text{defect}} = - \frac{1}{24\pi} \left(b \, \mathcal{R}_\Sigma + d_1\,K^\mu_{ab}K_\mu^{ab} - d_2 \, W_{ab}{}^{ab} \right),
\end{equation}
where $\mathcal{R}_\Sigma$ is the 2d Euler density, $K^\mu_{ab}$ is the traceless extrinsic curvature, and $W_{ab}{}^{ab}$ is the trace of the induced Weyl tensor on the surface $\Sigma$. The coefficient $b$ satisfies a monotonicity theorem \cite{Jensen:2015swa} and is determined by the logarithmic divergence of the defect free energy via the relation
\begin{equation}
\mathcal{F}_{\text{univ}} = -\frac{b}{3} \log(\mu R)\,,
\end{equation}
where $\mu$ is the renormalization scale and $R$ is the radius of the sphere.

The computation follows the one in \cite{Giombi:2023dqs}, but with generic $\Delta_{\phi}$. Summing up the three diagrams shown in Appendix \ref{g-fun} gives the following logarithmic term in the free energy
\begin{equation}
\mathcal{F}_{\text{univ}} = \left(\frac{N (1+\kappa)\hat{\varepsilon} \,h^2}{16 \pi ^2}-\frac{Nh^3}{12 \pi ^3}-\frac{N (N+2) \lambda h^2}{384 \pi ^4}\right) \log \left(\mu R\right)\,.
\end{equation}
This expression is very similar to the local case \cite{Trepanier:2023tvb,Giombi:2023dqs,Raviv-Moshe:2023yvq} apart from the $\kappa$ dependence in the first term. At the defect fixed point we can compute the Weyl anomaly coefficient $b$ finding
\begin{equation}
b_{\text{IR}} = - \frac{(8+4\kappa+N(1-\kappa))^3}{64(N+8)^3} \hat{\varepsilon}^3 + \text{O}\left(\hat{\varepsilon}^4\right) < b_{\text{UV}} = 0\,,
\end{equation}
which is consistent with the defect $b$-theorem \cite{Jensen:2015swa}.

 \section{Semiclassical construction of defects} \label{semiclassics}
In the previous sections, defects for the long-range $O(N)$ model have been realized as fixed points of defect RG flows triggered by interactions that are classically marginal. However, strongly relevant interactions might as well originate a defect RG flow that ends in an IR fixed point. The latter case is in general more difficult to study, since the defect coupling is usually not perturbatively small. One way to get around this problem is to look for non-trivial saddle points of the path integral that respects the defect-conformal symmetry group. Perturbation theory around these new saddles can then be used to compute quantum corrections for various observables.

\subsection{A warm-up example}

Let us illustrate this idea with a simple example. Consider the local $O(N)$ model, whose action is
\begin{equation}
S_{\text{bulk}} = \int d^d x \, \left(\frac{1}{2} \left( \partial \phi_a \right)^2 + \frac{\lambda_0}{4!} (\phi_a)^2\right)\,.
\end{equation}
We can add an interaction localized on a $p$-dimensional defect
\begin{equation}
S_{D}= h_0 \int d^p \tau \  \phi_1\,.
\end{equation}
When $d=4-\hat{\varepsilon}$ and $p=1$, the interaction is marginally relevant and one can therefore use standard diagrammatic techniques to compute observables \cite{Allais:2014fqa, Cuomo:2021kfm}. In the more general case, the beta function for the defect coupling reads
\begin{equation}\label{betah}
\beta_h= -\left(p-\Delta_\phi \right)h + \text{O}(\lambda) \,,
\end{equation}
with $\Delta_\phi = (d-2)/2$. If $d=4-\hat{\varepsilon}$ (but with $p$ generic), the bulk coupling constant undergoes a ``short'' RG flow from the Gaussian fixed point to the Wilson-Fisher fixed point. At any scale $\mu$ the following inequality holds
\begin{equation}
0 \leq \lambda(\mu) \leq \lambda_* = \frac{48\pi^2}{N+8}\,\hat{\varepsilon}+\text{O}(\hat{\varepsilon}^2)\,,
\end{equation}
and in the IR limit we have $\lim_{\mu \rightarrow 0} \lambda(\mu)= \lambda_*$. Using this bound, we can solve \eqref{betah} for $h(\mu)$
\begin{equation}
h(\mu)= h(\mu_0) \left( \frac{\mu}{\mu_0}\right)^{-(p-\Delta_\phi)+\text{O}(\hat{\varepsilon})}\,.
\end{equation}
It is now evident that if $p-\Delta_\phi \gg \hat{\varepsilon}$, then $h(\mu)$ rapidly flows to infinity in the IR limit, while $\lambda(\mu)$ is still perturbatively small. This new phase of the theory should correspond to a non-trivial saddle point of the path integral. The saddle point is described by a classical profile $\phi^a_{\text{cl}}(x)$ that satisfies the classical equations of motion. We can formulate an \emph{ansatz} that respects all the symmetries of the problem
\begin{equation}\label{ansatzsemiclass}
\phi^a_{\text{cl}}(x) = \delta_{a1} \frac{ \mathcal{N}_\phi\,a_{\text{cl}}}{|x_\perp|^{\Delta_{\text{cl}}}}\,.
\end{equation}
The equations of motion yield
\begin{equation}\label{semiclassiccalc}
\Box \phi^a_{\text{cl}}(x)=  \frac{\mathcal{N}_\phi\,a_{\text{cl}}}{|x_\perp|^{\Delta_{\text{cl}}+2}}\left( \Delta_{\text{cl}}\left(\Delta_{\text{cl}}+p+\varepsilon-2 \right)\right) \delta_{a1}=\frac{\lambda_0}{3!} \frac{{\mathcal{N}_\phi^{\,3}\,a_{\text{cl}}}^3}{|x_\perp|^{3\Delta_{\text{cl}}}}\,\delta_{a1}\,.
\end{equation}
This is readily solved
\begin{equation}\label{classicalresult}
\Delta_{\text{cl}}=1\,, \quad  \mathcal{N}_\phi^{\,2}\,{a_{\text{cl}}}^2=\frac{6}{\lambda_0}\left(p+\hat{\varepsilon}-1 \right)\,.
\end{equation}
The two solutions for $a_\text{cl}$ are associated to the $\mathbb{Z}_2$ symmetry $\phi_a\to -\phi_a$, which is explicitly broken by the defect. In the following, we focus on the solution $a_\text{cl}>0$.
To compute observables, one can expand the bulk action around this new saddle point
\begin{equation}
\phi^a(x)= \phi^a_{\text{cl}}(x)+\delta \phi^a(x)\,, \quad S'[ \delta \phi^a] = S_{\text{bulk}}[  \phi^a_{\text{cl}}+\delta \phi^a]-S_{\text{bulk}}[  \phi^a_{\text{cl}}]\,,
\end{equation}
and then use standard diagrammatic techniques with the new action $S'[ \delta \phi^a]$.\footnote{
In this realization, the presence of the defect is imposed trough the \emph{ansatz} \eqref{ansatzsemiclass} instead of an explicit defect term in the action.
} Note that $\phi^a_{\text{cl}}(x)$ plays the role of a classical external source in this new action.

For some simple observables, the semiclassical analysis is sufficient to determine their value at the IR fixed point at leading order. For instance, for the order parameter we have
\begin{equation}
\langle \phi^a (x) \rangle = \phi^a_{\text{cl}}(x) + \langle \delta \phi^a (x) \rangle\,,
\end{equation}
where the second term at the IR fixed point is of order $\hat{\varepsilon}^\frac{1}{2}$ and is therefore subleading.
Moreover, conformal symmetry imposes
\begin{equation}\label{defectonepoint}
\langle \phi^a (x) \rangle =  \delta_{a1} \frac{\mathcal{N}_\phi\,a_\phi}{|x_\perp|^{\Delta_\phi}}\,.
\end{equation}
From this it follows that at the IR fixed point
\begin{equation}
{a_\phi}^2= {a_{\text{cl}}}^2 (1+\text{O}(\hat{\varepsilon})) = \frac{24 \pi^2}{\lambda_*}\left(p-1 \right)+\text{O}(\hat{\varepsilon}^{\,0})\,.
\end{equation}
For the boundary case $p=3-\hat{\varepsilon}$, this is indeed the correct value of the coefficient of the one-point function \cite{Shpot:2019iwk}.

Note that for $p=1$ the classical contribution is vanishing as expected, since in that case the defect interaction is classically marginal.
Interestingly, the equations of motion toghether with conformal symmetry can still be used to find the coefficient of the one-point function at leading order. Indeed, from the original action $S$ it is evident that $\langle \phi^a \phi^b \phi^b \rangle = \langle \phi^a \rangle \langle \phi^b \rangle \langle \phi^b \rangle + \text{O}(\lambda)$.  Assuming the form of the one-point function at the fixed point \eqref{defectonepoint}, the expectation value of the equation of motion then gives\,\footnote{
Note that this is not equivalent as taking $p=1$ in \eqref{classicalresult} and expanding. Indeed to get the correct result one need $\Delta_\phi$ instead of $\Delta_{\text{cl}}$, toghether with the fact that at order $O(\varepsilon)$ the operator $\phi$ has no anomalous dimension.
}
\begin{equation}
{a_\phi}^2= \frac{N+8}{4}+\text{O}(\hat{\varepsilon})\,,
\end{equation}
which agrees with the results of \cite{Allais:2014fqa, Cuomo:2021kfm} computed with Feynman diagrams.

\subsection{The long-range $O(N)$ model}
Now let us consider the long-range $O(N)$ model with a localized magnetic field on a line. Recall that the action is
\begin{equation}
S= \int d^d x \, \left(\frac{1}{2}\phi_a \mathcal{L}_{\sigma} \phi_a + \frac{\lambda_0}{4!} (\phi_a)^2\right) +  h_0 \int  d \tau \  \phi_1\,. 
\end{equation}
With $\sigma=(d+\varepsilon)/2$, the bulk interaction is weakly relevant. However, in this case $\Delta_\phi=(d- \varepsilon)/4$, and the defect interaction is strongly relevant for $2<d<4$. Therefore, standard perturbation theory around the trivial vacuum fails. Instead, we need to carry out an analysis similar to the one of last section.
\subsubsection{Semiclassics} 
The first step to study this model is to look for non-trivial saddle points of the path integral of the form
\begin{equation}
\phi^a_{\text{cl}}(x) = \delta_{a1} \frac{\mathcal{N}_\phi\,a_{\text{cl}}}{|x_\perp|^{\Delta_{\text{cl}}}}\,.
\end{equation}
The non-local equations of motion are\begin{equation}\label{semiclassiccalcnonlooc}
 -\mathcal{L}_\sigma \phi^a_{\text{cl}}(x)=   -\frac{2^{\frac{d+ \varepsilon }{2}} \Gamma \!\left(\frac{d-1-\Delta_{\text{cl}}}{2} \right) \Gamma \!\left(\frac{d+\varepsilon+2\Delta_{\text{cl}}}{4} \right) \,\mathcal{N}_\phi\, a_{\text{cl}}}{\Gamma \!\left(\frac{\Delta_{\text{cl}}}{2}\right) \Gamma \!\left(\frac{d-2-\varepsilon-2\Delta_{\text{cl}}}{4}\right) |x_\perp|^{\Delta_{\text{cl}}+\frac{d+\varepsilon}{2}}}\,\delta_{a1} =\frac{\lambda_0}{3!} \frac{{\mathcal{N}_\phi^{\,3}\,a_{\text{cl}}}^3}{|x_\perp|^{3\Delta_{\text{cl}}}}\,\delta_{a1}\,,
\end{equation}
and they are solved by
\begin{equation}\label{deltaaclass}
\Delta_{\text{cl}}= \frac{d+\varepsilon}{4}\,, \quad  \mathcal{N}_\phi^{\,2}\,{a_{\text{cl}}}^2=- \frac{6}{\lambda_0} \,  \frac{2^{\frac{d+\varepsilon }{2}}\Gamma \!\left(\frac{3d-4-\varepsilon}{8} \right)\Gamma\! \left(\frac{3d+3\varepsilon}{8} \right) }{\Gamma \!\left(\frac{d-4-3\varepsilon}{8} \right)\Gamma \!\left(\frac{d+\varepsilon}{8} \right)}  \,,
\end{equation}
where as before we select the solution $a_\text{cl}>0$.
Reasoning as in the previous section, we can conclude that
\begin{equation}
\langle \phi^a (x) \rangle =  \delta_{a,1} \frac{\mathcal{N}_\phi\,a_\phi}{|x_\perp|^{\Delta_\phi}}\,.
\end{equation}
with
\begin{equation}\label{semiclassicallongrangedata}
\Delta_\phi= \frac{d-\varepsilon}{4}\,, \quad  {a_\phi}^2={a_{\text{cl}}}^2 (1+\text{O}(\varepsilon)) = - \frac{(N+8)\Gamma\!\left( \frac{3d-4}{4}\right)}{2^{\frac{d}{2}-1}\Gamma\!\left( \frac{d-4}{4}\right)\Gamma\!\left( \frac{d}{2}\right) \varepsilon}+\text{O}\left(\varepsilon^0\right)\,.
\end{equation}
Note that for $d=4+\text{O}(\varepsilon)$ the classical contribution vanishes, since we get perturbatively close to the trivial vacuum. 
As in the previous example, by considering the expectation value of the equation of motion, we can compute the $\text{O}(\hat{\varepsilon})$ contribution to the coefficient $a_\phi$ near $d=4$, where the defect becomes weakly coupled. This result can then be matched to the results of Section \ref{localizedmagneticfield}.

Let us briefly explain how the matching procedure works. In this section, we parameterize the defect conformal manifold (shown in Figure \ref{phases}) using the coordinates $(d, \varepsilon)$, near the crossover with GFF. In contrast, in Section \ref{epsexp}, we use the coordinates $(\hat{\varepsilon}, k)$ near the point $d=4$. To switch between these coordinate systems, we need to ensure that we are describing the same point $(d, \sigma)$. The corresponding condition is $d = 4 - \hat{\varepsilon}$ and $\varepsilon= \hat{\varepsilon}/k + \text{O}(\hat{\varepsilon}^2)$.

By considering the expectation value of the equation of motion and switching to the $(\hat{\varepsilon}, k)$ coordinates, we obtain:
\begin{equation}
a_\phi^2 = \frac{(N+8)(\kappa+1)}{8\, \kappa} + \text{O}(\hat{\varepsilon}),
\end{equation}
which precisely matches the results in \eqref{onepointclosetofour} and \eqref{onepointcoefficientclosetofour}.

\subsubsection{Quantum corrections} 
In order to compute quantum fluctuations around the new saddle point, we need to expand the action around it
\begin{equation}\label{semiclassicalaction}
\begin{split}
&S'[ \delta \phi^a] = S_{\text{bulk}}[  \phi^a_{\text{cl}}+\delta \phi^a]-S_{\text{bulk}}[  \phi^a_{\text{cl}}] = \int  d^d x\, \frac{1}{2}\bigg(   \Big( \delta^{ab} \mathcal{L}_\sigma
  \\
 & \quad \quad +\frac{ \lambda_0 \bar{\phi}^2}{6 |x_\perp|^{2\Delta_\text{cl}}}\left(\delta^{ab} +2 \delta^{a1}\delta^{b1}\right) \!\Big)\delta \phi_a \delta \phi_b + \frac{\lambda_0 \bar{\phi}}{6|x_\perp|^{\Delta_\text{cl}}} \, \delta \phi_a \delta \phi_a \delta 	\phi_1   +  \tfrac{\lambda_0}{4!}  (\delta \phi_a)^2 \bigg)\,.
\end{split}
\end{equation}
For convenience, we have defined $\bar{\phi}= \mathcal{N}_\phi \, a_\text{cl}$.
A similar analysis can also be done for a $O(N)$ breaking surface defect \cite{Giombi:2023dqs}.
In \eqref{semiclassicalaction}, there are new quadratic and cubic terms that break both Poincaré symmetry and $O(N)$ symmetry to $O(N-1)$, as expected. The quadratic term comes with coefficient $ \lambda_0 \bar{\phi}^2 $, which, by \eqref{deltaaclass}, is a pure number. One can either try to invert the quadratic part of the action or treat this new term as a perturbation. In the latter case, since the coefficient is not small, one must always consider an arbitrary number of insertions of this term. This is what we will do in the following, since the inversion of the quasratic operator turns out to be a very difficult task. On the other hand, since $ \lambda_0 \bar{\phi} \sim \sqrt{\lambda_0}$, the cubic term is perturbatively small.

As an instructive example, we will set up the computation of the leading-order quantum corrections to the order parameter  $\langle \delta\phi_a(x) \rangle$.
For concreteness, we focus on the case
$d=3$, which is the physically relevant one. 
The following analysis is similar to the one carried out for the extraordinary surface transition in \cite{Shpot:2019iwk}. However, our case is more complicated due to the fact that we have a line defect instead of a boundary.
The diagrams contributing to this one-point function at order $\sqrt{\lambda_0}$ all belong to a single family, which we can represent as follows
\begin{equation}\label{semiclassicaldiagram}
 \begin{tikzpicture}[scale=0.4, baseline]
       \draw[thick, black] (-4,0)--(0,0);
	  	   \node[above] at (-4.1,0) {$\delta\phi_a$};
	  	    \draw[fill= white] (-2,0) circle (15pt);	
 \draw[fill= white, draw=black, line width=0.25mm, pattern=north east lines] (-2,0) circle (15pt);
 \draw[fill= white, draw=black, line width=0.25mm] (1.5,0) circle (42pt);
  \draw[black, fill=black] (0,0) circle (4pt);
   \draw[black, fill=black] (2.4,1.15) circle (4pt);
    \draw[fill= white](2.4,-1.15) circle (15pt);	
   \draw[fill= white, draw=black, line width=0.25mm, pattern=north east lines] (2.4,-1.15) circle (15pt);	
	\end{tikzpicture} 
\end{equation}
where the dots represent the quadratic and cubic interactions, and the ``bubble'' is the dressed propagator defined by
\begin{equation}\label{dressedpropagator}
 \begin{tikzpicture}[scale=0.4, baseline=-0.1cm]
       \draw[thick, black] (-4,0)--(0,0);
	  	    \draw[fill= white] (-2,0) circle (15pt);	
 \draw[fill= white, draw=black, line width=0.25mm, pattern=north east lines] (-2,0) circle (15pt);
	\end{tikzpicture} 
	\hspace{0.5cm} = \hspace{0.5cm} 
	 \begin{tikzpicture}[scale=0.4, baseline=-0.1cm]
       \draw[thick, black] (-3,0)--(0,0);
	\end{tikzpicture} 
	 	\hspace{0.4cm} + \hspace{0.4cm} 
	 		\begin{tikzpicture}[scale=0.4, baseline=-0.1cm]
       \draw[thick, black] (-3,0)--(0,0);
    \draw[black, fill=black] (-1.5,0) circle (4pt);	
	\end{tikzpicture} 
	\hspace{0.4cm} + \hspace{0.4cm} 
	 		\begin{tikzpicture}[scale=0.4, baseline=-0.1cm]
       \draw[thick, black] (-3,0)--(0,0);
    \draw[black, fill=black] (-1,0) circle (4pt);
    \draw[black, fill=black] (-2,0) circle (4pt);	
	\end{tikzpicture} 
	\hspace{0.4cm} + \ \, \ldots 
\end{equation}
In \eqref{semiclassicaldiagram} there is always at least one quadratic interaction inserted in the loop to avoid tadpole contributions.
This diagram is already difficult to compute analytically, mainly because the terms in \eqref{dressedpropagator} involve challenging integrals, whose results must then be resummed.

We begin by extracting the divergent part of \eqref{semiclassicaldiagram}. It turns out that the divergence arises only from the following diagram
\begin{equation}\label{aphidivergencies}
 \begin{tikzpicture}[scale=0.4,baseline]
       \draw[thick, black] (-4,0)--(0,0);
	  	   \node[above] at (-4.1,0) {$\delta\phi_a$};
	  	    \draw[fill= white] (-2,0) circle (15pt);	
 \draw[fill= white, draw=black, line width=0.25mm, pattern=north east lines] (-2,0) circle (15pt);
 \draw[fill= white, draw=black, line width=0.25mm] (1.5,0) circle (42pt);
  \draw[black, fill=black] (0,0) circle (4pt);
   \draw[black, fill=black] (2.95,0) circle (4pt);
	\end{tikzpicture} 
\end{equation}
This can be verified by checking all the other diagrams and observing that they do not exhibit logarithmic divergences.
To evaluate \eqref{aphidivergencies}, we use the following recursion relation
\begin{equation}\label{recursion}
 \begin{tikzpicture}[scale=0.4, baseline=-0.1cm]
       \draw[thick, black] (-4,0)--(0,0);
	  	    \draw[fill= white] (-2,0) circle (15pt);	
 \draw[fill= white, draw=black, line width=0.25mm, pattern=north east lines] (-2,0) circle (15pt);
 \draw[fill= white, draw=black, line width=0.25mm] (1.5,0) circle (42pt);
  \draw[black, fill=black] (0,0) circle (4pt);
   \draw[black, fill=black] (2.95,0) circle (4pt);
	\end{tikzpicture} 
		\hspace{0.4cm} = \hspace{0.4cm} 
		 \begin{tikzpicture}[scale=0.4, baseline=-0.1cm]
       \draw[thick, black] (-2,0)--(0,0);
 \draw[fill= white, draw=black, line width=0.25mm] (1.5,0) circle (42pt);
  \draw[black, fill=black] (0,0) circle (4pt);
   \draw[black, fill=black] (2.95,0) circle (4pt);
	\end{tikzpicture} 
	\hspace{0.4cm} + \hspace{0.4cm} 
	 \begin{tikzpicture}[scale=0.4, baseline=-0.1cm]
       \draw[thick, black] (-4.2,0)--(0,0);
	  	    \draw[fill= white] (-1.5,0) circle (15pt);	
 \draw[fill= white, draw=black, line width=0.25mm, pattern=north east lines] (-1.5,0) circle (15pt);
 \draw[fill= white, draw=black, line width=0.25mm] (1.5,0) circle (42pt);
  \draw[black, fill=black] (0,0) circle (4pt);
   \draw[black, fill=black] (2.95,0) circle (4pt);
   \draw[black, fill=black] (-3,0) circle (4pt);
	\end{tikzpicture} 
	 \ \,.
\end{equation}
The first diagram on the right-hand side can be easily computed
\begin{align}
& \begin{tikzpicture}[scale=0.4, baseline=-0.1cm]
       \draw[thick, black] (-2,0)--(0,0);
 \draw[fill= white, draw=black, line width=0.25mm] (1.5,0) circle (42pt);
  \draw[black, fill=black] (0,0) circle (4pt);
   \draw[black, fill=black] (2.95,0) circle (4pt);
	\end{tikzpicture} 
	\hspace{0.2cm} =   \ \delta_{a1}\, \frac{A_0}{|x_\perp|^\frac{3-3\varepsilon}{4}}\,, \nonumber \\
	& A_0 = \frac{\sqrt{\lambda_0 }(N+8) \Gamma \!\left( \frac{1}{4}\right)}{2^\frac{15}{4}\sqrt{3}\,\pi^\frac{5}{2}} \bigg( \frac{1}{\varepsilon} \, + \\
	& \quad \quad + \frac{\left(48-2 \pi +3 \gamma_E +10 \sqrt{2} \log \left(2-\sqrt{2}\right)+\left(13-5 \sqrt{2}\right) \log (2)\right)}{4} +\text{O}(\varepsilon) \bigg)\,. \nonumber
\end{align}
Using dimensional analysis, we can make the following ansatz for the left-hand side
\begin{equation}\label{ansatzonept}
\begin{tikzpicture}[scale=0.4, baseline=-0.1cm]
       \draw[thick, black] (-4,0)--(0,0);
	  	    \draw[fill= white] (-2,0) circle (15pt);	
 \draw[fill= white, draw=black, line width=0.25mm, pattern=north east lines] (-2,0) circle (15pt);
 \draw[fill= white, draw=black, line width=0.25mm] (1.5,0) circle (42pt);
  \draw[black, fill=black] (0,0) circle (4pt);
   \draw[black, fill=black] (2.95,0) circle (4pt);
	\end{tikzpicture} 
	\hspace{0.2cm} =  \ \delta_{a1}\, \frac{A}{|x_\perp|^\frac{3-3\varepsilon}{4}}\,.
\end{equation}
Substituting this ansatz into the recursion relation in \eqref{recursion}, and evaluating the integral in the last term on the right-hand side, yields a linear equation that can be solved for $A$
\begin{align}\label{Acoeff}
&A = \frac{\sqrt{\lambda_0 }(N+8) \Gamma \!\left( \frac{1}{4}\right)}{2^\frac{19}{4}\sqrt{3}\,\pi^\frac{5}{2}} \bigg( \frac{1}{\varepsilon} \, + \\
	& \quad \quad \quad  + \frac{32-2 \pi +3 \gamma_E +13 \log (2)-3 \sqrt{2} \log \left(2 \sqrt{2}+3\right)}{4} +\text{O}(\varepsilon) \bigg)\,.  \nonumber
\end{align}
As we argued, \eqref{Acoeff} captures the full divergence of the more complicated diagram in \eqref{semiclassicaldiagram}. A non-trivial consistency check is that this divergence exactly cancels the one arising from the coupling renormalization in the classical term. More specifically, we have
\begin{equation}\label{semiclassicalorderparameter}
\langle \phi_a(x) \rangle = \phi^a_\text{cl} + \langle \delta \phi_a(x) \rangle\,,
\end{equation}
where $ \phi^a_\text{cl} \sim 1/\sqrt{\lambda_0}$ as given by \eqref{deltaaclass}. Since the bulk renormalizes independently, we also have that $\lambda_0 = \mu^\varepsilon \lambda (1+\lambda(N+8)/(12 \pi^2 \varepsilon) + \text{O}(\lambda^2))$, according to \eqref{Zlambda}.
Therefore, the first term in \eqref{semiclassicaldiagram} has a divergence at order $\sqrt{\lambda}$, which is exactly canceled by the divergence in the second term that we just computed. This ensures that $\langle \phi_a (x) \rangle$ remains finite.
The finite term corresponds to the order $\varepsilon^0$ correction to \eqref{semiclassicallongrangedata}. Part of this finite term is given in \eqref{Acoeff}, while the rest comes from diagrams that contain more than one quadratic interaction in the loop. Their contributions can be collected into one single diagram
\begin{equation}\label{finitetermdiagrams}
 \begin{tikzpicture}[scale=0.4, baseline]
       \draw[thick, black] (-4,0)--(0,0);
	  	    \draw[fill= white] (-2,0) circle (15pt);	
 \draw[fill= white, draw=black, line width=0.25mm, pattern=north east lines] (-2,0) circle (15pt);
 \draw[fill= white, draw=black, line width=0.25mm] (1.5,0) circle (42pt);
  \draw[black, fill=black] (0,0) circle (4pt);
 \draw[black, fill=black] (1.6,-1.45) circle (4pt);
      \draw[black, fill=black] (1.6,1.45) circle (4pt);
    \draw[fill= white](2.95,0) circle (15pt);	
   \draw[fill= white, draw=black, line width=0.25mm, pattern=north east lines] (2.95,0) circle (15pt);	
	\end{tikzpicture} 
\end{equation}
Furthermore, using a recursion similar to the one in \eqref{recursion}, it suffices to compute the following 
\begin{equation}\label{finitetermdiagrams2}
 \begin{tikzpicture}[scale=0.4, baseline=-0.1cm]
       \draw[thick, black] (-2,0)--(0,0);
 \draw[fill= white, draw=black, line width=0.25mm] (1.5,0) circle (42pt);
  \draw[black, fill=black] (0,0) circle (4pt);
 \draw[black, fill=black] (1.6,-1.45) circle (4pt);
      \draw[black, fill=black] (1.6,1.45) circle (4pt);
    \draw[fill= white](2.95,0) circle (15pt);	
   \draw[fill= white, draw=black, line width=0.25mm, pattern=north east lines] (2.95,0) circle (15pt);	
	\end{tikzpicture} 
	\hspace{0.4cm} = \hspace{0.4cm} 
	 \begin{tikzpicture}[scale=0.4, baseline=-0.1cm]
       \draw[thick, black] (-2,0)--(0,0);
 \draw[fill= white, draw=black, line width=0.25mm] (1.5,0) circle (42pt);
  \draw[black, fill=black] (0,0) circle (4pt);
   \draw[black, fill=black] (2.4,1.15) circle (4pt);
     \draw[black, fill=black] (2.4,-1.15) circle (4pt);
	\end{tikzpicture} 
	\hspace{0.4cm} + \hspace{0.4cm} 
		 \begin{tikzpicture}[scale=0.4, baseline=-0.1cm]
       \draw[thick, black] (-2,0)--(0,0);
 \draw[fill= white, draw=black, line width=0.25mm] (1.5,0) circle (42pt);
  \draw[black, fill=black] (0,0) circle (4pt);
 \draw[black, fill=black] (1.6,-1.45) circle (4pt);
      \draw[black, fill=black] (1.6,1.45) circle (4pt);
      \draw[black, fill=black] (2.95,0) circle (4pt);
	\end{tikzpicture} 
	\hspace{0.4cm} + \ \, \ldots
\end{equation}
Unfortunately, performing this computation analytically is difficult. However, all the diagrams in the sum in \eqref{finitetermdiagrams2} are finite for $\varepsilon=0$ and can be evaluated numerically. The result of the sum can then be extrapolated using standard numerical techniques, such as Padé approximation. We leave this analysis for future investigations.

\section*{Acknowledgments} 
We would like to thank Adam Chalabi, Gabriel Cuomo, Charlotte Kristjansen, Marco Meineri, Edoardo Lauria and Miguel Paulos for useful discussions. EdS's research is partially supported by the MUR PRIN contract 2020KR4KN2 “String
Theory as a bridge between Gauge Theories and Quantum Gravity” and by the INFN project ST\&FI “String Theory \& Fundamental Interactions”.

\appendix

\section{Diagrams for the defect coupling renormalization} \label{locmagdiag}
In this appendix, we list some of the diagrams that are relevant to the renormalization of defect couplings $h$ in section \ref{epsexp}. The first set of diagrams relates to the case of the localized magnetic field. Recall the notation $w_A^{(d)} = (4\pi)^{d/2} 2^{-A} \Gamma\left(\frac{d-A}{2}\right)/\Gamma\left(\frac{A}{2}\right)$

 \begin{equation}
  \begin{aligned}
      \begin{tikzpicture}[scale=0.4, baseline]
       \draw[thick, black] (0,3)--(0,0);
	 \draw[double,thick,blue] (-2,0)--(2,0);
	  	  \draw[blue, fill=blue]   (0,0) circle (4pt);
	  	   \node[above] at (0,3) {$\phi$};
	\end{tikzpicture} = -\mathcal{N}_\phi^2 h_0 \tfrac{\sqrt{\pi} \Gamma\left(\Delta_\phi-\tfrac{1}{2}\right)}{\Gamma(\Delta_\phi)} \tfrac{1}{|x_\perp|^{2\Delta_\phi - 1}}\,,
 \end{aligned}
  \end{equation}

  \begin{equation}
      \begin{tikzpicture}[scale=0.4, baseline]
       \draw[thick, black] (0,3)--(0,0);
       \draw[double,thick,blue] (-3,0)--(3,0);
	  	  \draw[blue, fill=blue]   (0,0) circle (4pt);
      \draw[blue, fill=black]   (0,1.5) circle (4pt);
       \draw[thick, black]    (0,1.5)--(-2,0);
       \draw[thick, black]    (0,1.5)--(2,0);
       \draw[blue, fill=blue]   (-2,0) circle (4pt);
      \draw[blue, fill=blue]   (2,0) circle (4pt);
	  	   \node[above] at (0,3) {$\phi$};
	\end{tikzpicture} = \tfrac{\left(\mathcal{N}_\phi^2\right)^4 \lambda_0 h_0^3}{6}  \left[\tfrac{\sqrt{\pi} \Gamma\left(\Delta_\phi - \tfrac{1}{2}\right)}{\Gamma(\Delta_\phi)}\right]^4 \tfrac{w_{2\Delta_\phi - 1}^{(d-1)} w_{6\Delta_\phi - 3}^{(d-1)}}{w_{8\Delta_\phi - 3 - d}^{(d-1)}} \tfrac{1}{|x_\perp|^{8\Delta_\phi - d - 3}}\,,
  \end{equation}

  \begin{equation}
      \begin{tikzpicture}[scale=0.4, baseline]
       \draw[thick, black] (0,5)--(0,0);
	 \draw[double,thick,blue] (-3,0)--(3,0);
	  	  \draw[blue, fill=blue]   (0,0) circle (4pt);
      \draw[blue, fill=black]   (0,3.5) circle (4pt);
      \draw[blue, fill=black]   (0,1.5) circle (4pt);
      \draw[thick, black]    (0,3.5) to[out=0,in=90] (1,2.5) to[out=270,in=0]  (0,1.5);
      \draw[thick, black]    (0,3.5) to[out=180,in=90] (-1,2.5) to[out=270,in=180]  (0,1.5);
	  \node[above] at (0,5) {$\phi$};
	\end{tikzpicture} = -\tfrac{(N+2) \left(\mathcal{N}_\phi^2\right)^5 \lambda_0^2 h_0}{18} \tfrac{\pi^{\frac{3}{2}} \Gamma\left(\Delta_\phi - \tfrac{1}{2}\right)^2 \Gamma\left(3\Delta_\phi - \tfrac{1}{2}\right)}{\Gamma(\Delta_\phi)^2 \Gamma(3\Delta_\phi)} \tfrac{w_{6\Delta_\phi - 1}^{(d-1)} \left[w_{2\Delta_\phi - 1}^{(d-1)}\right]^2}{w_{10\Delta_\phi - 2d - 1}^{(d-1)}} \tfrac{1}{|x_\perp|^{10\Delta_\phi - 2d - 1}}\,,
  \end{equation}

    \begin{equation}
      \begin{tikzpicture}[scale=0.4, baseline]
       \draw[thick, black] (0,5)--(0,0);
	 \draw[double,thick,blue] (-4,0)--(4,0);
  \draw[blue, fill=blue]   (0,0) circle (4pt);
  \draw[thick, black]    (0,1.5)--(-1,0);
       \draw[thick, black]    (0,1.5)--(1,0);
       \draw[thick, black]    (0,3)--(-2,0);
       \draw[thick, black]    (0,3)--(2,0);
      \draw[blue, fill=blue]   (-2,0) circle (4pt);
      \draw[blue, fill=blue]   (2,0) circle (4pt);
      \draw[blue, fill=blue]   (-1,0) circle (4pt);
      \draw[blue, fill=blue]   (1,0) circle (4pt);
      \draw[blue, fill=black]   (0,1.5) circle (4pt);
      \draw[blue, fill=black] (0,3) circle (4pt);
	  	   \node[above] at (0,5) {$\phi$};
	\end{tikzpicture} = - \tfrac{\left(\mathcal{N}_\phi^2\right)^7 \lambda_0^2 h_0^5}{12} \left[\tfrac{\sqrt{\pi} \Gamma\left(\Delta_\phi - \tfrac{1}{2}\right)}{\Gamma(\Delta_\phi)}\right]^7 \tfrac{\left[w_{2\Delta_\phi - 1}^{(d-1)}\right]^2 w_{6\Delta_\phi - 3}^{(d-1)} w_{12\Delta_\phi - d - 5}^{(d-1)}}{w_{8\Delta_\phi - d - 3}^{(d-1)} w_{14\Delta_\phi - 2d - 5}^{(d-1)}} \tfrac{1}{|x_\perp|^{14\Delta_\phi - 2d - 5}}\,,
  \end{equation}

      \begin{equation}
      \begin{aligned}
      \begin{tikzpicture}[scale=0.3, baseline]
       \draw[thick,black] (-3,4)--(-2,2);
       \draw[thick,black] (-2,2)--(2,2);
       \draw[thick,black] (-2,2)--(-3,0);
       \draw[thick,black] (2,2)--(1,0);
       \draw[thick,black] (2,2)--(3,0);
       \draw[thick, black] (-2,2) to [out=90,in=180] (0,4) to [out=0,in=90] (2,2);
       \draw[blue, fill=black] (-2,2) circle (4pt);
       \draw[blue, fill=black] (2,2) circle (4pt);
	 \draw[double,thick,blue] (-4,0)--(4,0);
  \draw[blue, fill=blue] (-3,0) circle (4pt);
  \draw[blue, fill=blue] (3,0) circle (4pt);
  \draw[blue, fill=blue] (1,0) circle (4pt);
	  \node[above] at (-3,4) {$\phi$};
	\end{tikzpicture} &= - \tfrac{(N+8) \left(\mathcal{N}_\phi^2\right)^6 \lambda_0^2 h_0^3}{36} \tfrac{\Gamma\left(\Delta_\phi - \frac{1}{2}\right)^4 \pi^{\frac{5}{2}} \Gamma\left(2\Delta_\phi - \tfrac{1}{2}\right)}{\Gamma(\Delta_\phi)^4 \Gamma(2\Delta_\phi)} \\ & \times \tfrac{w^{(d-1)}_{4\Delta_\phi - 1} w^{(d-1)}_{4\Delta_\phi - 2} w^{(d-1)}_{2\Delta_\phi - 1} w^{(d-1)}_{10\Delta_\phi - d - 3}}{w^{(d-1)}_{8\Delta_\phi - d - 2} w^{(d-1)}_{12\Delta_\phi - 2d - 3}} \tfrac{1}{|x_\perp|^{12\Delta_\phi - 2d - 3}}\,.
	\end{aligned}
  \end{equation}

\noindent The second set of diagrams appears in the renormalization of the surface defect coupling
  \begin{equation}
\begin{aligned}
\begin{tikzpicture}[scale=0.4, baseline]
	 \draw[double,thick,blue] (-3,0)--(3,0);
	  	  \draw[blue, fill=blue]   (0,0) circle (4pt);
          \draw[blue, fill=black]   (0,5) circle (4pt);
      \draw[thick, black]    (0,5) to[out=-45,in=90] (1,2.5) to[out=270,in=45]  (0,0);
      \draw[thick, black]    (0,5) to[out=225,in=90] (-1,2.5) to[out=270,in=135]  (0,0);
	  \node[above] at (0.2,5) {$\phi^2$};
	\end{tikzpicture}
	&= -2h_0 \left(\mathcal{N}_\phi^2\right)^2 N \frac{\pi}{2\Delta_\phi - 1} \tfrac{1}{|x_\perp|^{4\Delta_\phi - 2}}\,,
\end{aligned}
\end{equation}

\begin{equation}
\begin{aligned}
 \begin{tikzpicture}[scale=0.4, baseline]
	 \draw[double,thick,blue] (-3,0)--(3,0);
     \draw[thick,black] (0,5)--(-1.7,0);
     \draw[thick,black] (0,5)--(1.7,0);
     \draw[thick,black] (-1.7,0) to[out=90,in=180] (0,1) to[out=0,in=90] (1.7,0);
	  	  \draw[blue, fill=blue]   (-1.7,0) circle (4pt);
           \draw[blue, fill=blue]   (1.7,0) circle (4pt);
          \draw[blue, fill=black]   (0,5) circle (4pt);
	  \node[above] at (0.2,5) {$\phi^2$};
	\end{tikzpicture}
	&= 4 \left(\mathcal{N}_\phi^2\right)^3N h_0^2 \tfrac{\pi^3 \Gamma\left(2\Delta_\phi - 1\right)^2 \Gamma\left(3\Delta_\phi - 2\right)}{\Gamma(\Delta_\phi)^3 \Gamma\left(4\Delta_\phi - 2\right) \sin(\pi \Delta_\phi)} \tfrac{1}{|x_\perp|^{6\Delta_\phi - 4}}\,,
\end{aligned}
\end{equation}

\begin{equation}
\begin{aligned}
\begin{tikzpicture}[scale=0.4, baseline]
	 \draw[double,thick,blue] (-3,0)--(3,0);
	  	  \draw[blue, fill=blue]   (0,0) circle (4pt);
          \draw[blue, fill=black]   (0,5) circle (4pt);
          \draw[blue, fill=black]   (0,2.5) circle (4pt);
      \draw[thick, black]    (0,5) to[out=0,in=90] (1,3.75) to[out=270,in=0]  (0,2.5);
      \draw[thick, black]    (0,2.5) to[out=0,in=90] (1,1.25) to[out=270,in=0]  (0,0);
      \draw[thick, black]    (0,2.5) to[out=180,in=90] (-1,1.25) to[out=270,in=180]  (0,0);
      \draw[thick, black]    (0,5) to[out=180,in=90] (-1,3.75) to[out=270,in=180]  (0,2.5);
	  \node[above] at (0.2,5) {$\phi^2$};
	\end{tikzpicture} &= \tfrac{h_0 \lambda_0 \left(\mathcal{N}_\phi^2\right)^4 N(N+2)}{3} \tfrac{\left[w_{4\Delta_\phi}^{(d)}\right]^2}{w_{8\Delta_\phi - d}^{(d)}} \tfrac{\pi}{4\Delta_\phi - 1 - \tfrac{d}{2}} \tfrac{1}{|x_\perp|^{8\Delta_\phi - d - 2}}\,.
\end{aligned}
\end{equation}

\section{Useful integrals}

The diagrams above make use of the following integrals.

\paragraph{Integral over a bulk vertex}

\begin{equation}
I(x) := \int \frac{d^dy}{|x-y|^A |y|^B} = \frac{w_A^{(d)} w_B^{(d)}}{w_{A+B-d}^{(d)}} \frac{1}{|x|^{A+B-d}}
\label{bulk vertex}
\end{equation}
\paragraph{Integral over a defect vertex}
\begin{equation}
\begin{aligned}
\int \frac{d^p\tau}{|x-x(\tau)|^{\alpha}} &= \frac{\pi^\frac{p}{2}\Gamma\left(\frac{\alpha-p}{2}\right)}{\Gamma\left(\frac{\alpha}{2}\right)}\frac{1}{|x_{\perp}|^{\alpha-p}}\,.
\end{aligned}
\end{equation}
\paragraph{Integral over a bulk vertex with defect propagator}
\begin{equation}
\begin{aligned}
\int\frac{d^dy}{\left|x-y\right|^\alpha\left|y_{\|}\right|^\beta} &=\frac{\pi^\frac{d-p}{2}\Gamma\left(\frac{\alpha-d+p}{2}\right)}{\Gamma\left(\frac{\alpha}{2}\right)}\frac{w_{\alpha-d+p}^{(p)}w_{\beta}^{(p)}}{w_{\alpha+\beta-d}^{(p)}}\frac{1}{\left|x_{\|}\right|^{\alpha+\beta-d}}\,.
\end{aligned}
\end{equation}
\paragraph{Three propagators}
The following integral is used in the renormalization of the surface defect coupling, and is particularly challenging to compute by hand (borrowing notation from \cite{Giombi:2023dqs}, see \cite{Smirnov:2006ry} for a derivation). One way to proceed is to use Schwinger parametrization and the inversion formula for the gamma function
\begin{equation}
\begin{aligned}
F &= \int\frac{d^dkd^dl}{\left(k^2+m^2\right)^{\lambda_1}\left[\left(k+l\right)^2\right]^{\lambda_2}\left(l^2+m^2\right)^{\lambda_3}}
\\ &= \frac{\pi^d\Gamma\left(\lambda_1+\lambda_2-\frac{d}{2}\right)\Gamma\left(\lambda_2+\lambda_3-\frac{d}{2}\right)\Gamma\left(\frac{d}{2}-\lambda_2\right)\Gamma\left(\lambda_1+\lambda_2+\lambda_3-d\right)}{\Gamma\left(\lambda_1\right)\Gamma\left(\lambda_3\right)\Gamma\left(\lambda_1+2\lambda_2+\lambda_3-d\right)\Gamma\left(\frac{d}{2}\right)}\frac{1}{\left(m^2\right)^{\lambda_1+\lambda_2+\lambda_3-d}}\,.
\end{aligned}
\end{equation}

\section{Diagrams for the $g$-function close to four dimensions} \label{g-fun}

First, we list the diagrams useful to the computation of the $g$-function close to four dimensions for the localized magnetic field. The second diagram is challenging to compute exactly, and the result up to $\mathcal{O}(\hat{\varepsilon})$ given below makes use of the integral \eqref{challenging}.

\begin{equation}
\begin{tikzpicture}[scale=0.4,baseline]
\draw[double,thick,blue] (0,3) to[out=0,in=90] (3,0) to[out=270,in=360] (0,-3) to[out=180,in=270] (-3,0) to[out=90,in=180] (0,3);
\draw[thick,black] (-3,0)--(3,0);
\draw[blue,fill=blue] (3,0) circle (4pt);
\draw[blue,fill=blue] (-3,0) circle (4pt);
\end{tikzpicture}= \mathcal{N}_\phi^2 2^{1 - 2\Delta_\phi} h_0^2 R^{2-2\Delta_\phi} \pi \frac{\Gamma\left(\frac{1}{2} - \Delta_\phi\right) \sqrt{\pi}}{ \Gamma\left(1 - \Delta_\phi\right)}\,,
\end{equation}

\begin{equation}
\begin{tikzpicture}[scale=0.4,baseline]
\draw[double,thick,blue] (0,3) to[out=0,in=90] (3,0) to[out=270,in=360] (0,-3) to[out=180,in=270] (-3,0) to[out=90,in=180] (0,3);
\draw[thick,black] (0,0)--(3*0.707,3*0.707);
\draw[thick,black] (0,0)--(-3*0.707,3*0.707);
\draw[thick,black] (0,0)--(3*0.707,-3*0.707);
\draw[thick,black] (0,0)--(-3*0.707,-3*0.707);
\draw[blue,fill=black] (0,0) circle (4pt);
\draw[blue,fill=blue] (3*0.707,3*0.707) circle (4pt);
\draw[blue,fill=blue] (-3*0.707,3*0.707) circle (4pt);
\draw[blue,fill=blue] (3*0.707,-3*0.707) circle (4pt);
\draw[blue,fill=blue] (-3*0.707,-3*0.707) circle (4pt);
\end{tikzpicture} = - \frac{\lambda_0 h_0^4}{384 \pi^2} + \text{O}(\hat{\varepsilon})\,.
\label{challenging diagram}
\end{equation}

\noindent Second, the free energy for the surface defect is computed using the diagrams below

\begin{equation}
\begin{aligned}
\begin{tikzpicture}[scale=0.4,baseline]
\draw[double,thick,blue] (0,3) to[out=0,in=90] (3,0) to[out=270,in=360] (0,-3) to[out=180,in=270] (-3,0) to[out=90,in=180] (0,3);
\draw[thick,black] (-3,0) to[out=90,in=180] (0,1.5) to[out=0,in=90] (3,0);
\draw[thick,black] (-3,0) to[out=270,in=180] (0,-1.5) to[out=360,in=270] (3,0);
\draw[blue,fill=blue] (3,0) circle (4pt);
\draw[blue,fill=blue] (-3,0) circle (4pt);
\end{tikzpicture} &= h_0^2 \left(\mathcal{N}_\phi^2\right)^2 N \oint \frac{d^2x d^2y}{|x-y|^{4\Delta_\phi}} = -h_0^2 \left(\mathcal{N}_\phi^2\right)^2 N \frac{(2R)^{4(1 - \Delta_\phi)} \pi^2 }{2\Delta_\phi-1}\,,
\end{aligned}
\end{equation}

\begin{equation}
\begin{aligned}
\begin{tikzpicture}[scale=0.4,baseline]
\draw[double,thick,blue] (0,3) to[out=0,in=90] (3,0) to[out=270,in=360] (0,-3) to[out=180,in=270] (-3,0) to[out=90,in=180] (0,3);
\draw[thick, black] (-3*0.866,-3*0.5)--(3*0.866,-3*0.5);
\draw[thick, black] (-3*0.866,-3*0.5)--(0,3);
\draw[thick, black] (3*0.866,-3*0.5)--(0,3);
\draw[blue,fill=blue] (3*0.866,-3*0.5) circle (4pt);
\draw[blue,fill=blue] (-3*0.866,-3*0.5) circle (4pt);
\draw[blue,fill=blue] (0,3) circle (4pt);
\end{tikzpicture}
&= - h_0^3 \left(\mathcal{N}_\phi^2\right)^3 \frac{4N}{3} R^{6(1-\Delta_\phi)} \frac{8\pi^{\frac{9}{2}} \Gamma\left(2 - 3\Delta_\phi\right)}{\Gamma\left(\frac{3}{2} - \Delta_\phi\right)^3}\,,
\end{aligned}
\end{equation}

\begin{equation}
\begin{aligned}
\begin{tikzpicture}[scale=0.4,baseline]
\draw[double,thick,blue] (0,3) to[out=0,in=90] (3,0) to[out=270,in=360] (0,-3) to[out=180,in=270] (-3,0) to[out=90,in=180] (0,3);
\draw[thick,black] (-3,0) to[out=90,in=180] (-1.5,1.5) to[out=0,in=90] (0,0);
\draw[thick,black] (0,0) to[out=90,in=180] (1.5,1.5) to[out=0,in=90] (3,0);
\draw[thick,black] (-3,0) to[out=270,in=180] (-1.5,-1.5) to[out=360,in=270] (0,0);
\draw[thick,black] (0,0) to[out=270,in=180] (1.5,-1.5) to[out=360,in=270] (3,0);
\draw[blue,fill=blue] (3,0) circle (4pt);
\draw[blue,fill=blue] (-3,0) circle (4pt);
\end{tikzpicture}
&= - \lambda_0 h_0^2 \left(\mathcal{N}_\phi^2\right)^4 \frac{N(N+2)}{6} \frac{\left[w_{4\Delta_\phi}^{(d)}\right]^2}{w_{8\Delta_\phi - d}^{(d)}} \frac{(2R)^{d + 4 - 8\Delta_\phi} \pi^2 }{\frac{d}{2} + 1-4\Delta_\phi}\,.
\end{aligned}
\end{equation}

\subsection{Useful integrals for the computation of $g$-functions}

The diagrams above make use of the following integrals.

\paragraph{Circular integral over one angle}
\begin{equation}
\begin{aligned}
\oint \frac{d\tau}{|x - x(\tau)|^{2\Delta}} &= \int_{-\pi}^{\pi} \frac{Rd\theta}{\left(\left|x_{\|}\right|^2+ \left|x_{\perp}\right|^2 + R^2-2R \left|x_{\|}\right|\cos{\theta}\right)^{\Delta}} 
\\ &= \frac{2\pi R}{\left[\left(\left|x_{\|}\right| + R\right)^2 + \left|x_{\perp}\right|^2\right]^\Delta} {}_2F_1\left(\frac{1}{2},\Delta;1,\frac{4R\left|x_{\|}\right| }{\left(\left|x_{\|}\right|  + R\right)^2 + \left|x_{\perp}\right|^2}\right)\,,
\label{hypergeometric}
\end{aligned}
\end{equation}
 where we've used the integral representation of the hypergeometric function ${}_2F_1
$. 
\paragraph{Circular integral over two angles}
\begin{equation}
\begin{aligned}
\oint  \frac{d\tau_1 d\tau_2}{|x(\tau_1) - x(\tau_2)|^{2\Delta}}&= 4^{1 - \Delta} R^{2-2\Delta} \pi \frac{\Gamma\left(\frac{1}{2} - \Delta\right) \sqrt{\pi}}{ \Gamma\left(1 - \Delta\right)}\,.
\end{aligned}
\end{equation}
\paragraph{Second diagram in the $g$-function for the localized magnetic field}

The computation of diagram \eqref{challenging diagram} is intricate. Calculating it amounts to evaluating the following quantity
\begin{equation}
\begin{aligned}
 - \frac{\lambda_0 h_0^4}{4!} \left(\mathcal{N}_\phi^2\right)^4 \int d^dx \left(\oint \frac{d\tau}{|x-x(\tau)|^{2\Delta_\phi}}\right)^4 = - \frac{\lambda_0 h_0^4}{4!} I\,.
\end{aligned}
\end{equation}
 The integral over the circular defect can be computed using equation \eqref{hypergeometric}. It simplifies to
\begin{equation}
\begin{aligned}
I = 2 \pi \left(2\pi R\right)^4 \left(\mathcal{N}_\phi^2\right)^4 S_{2 - \varepsilon} \int_0^\infty\!\! dr \int_{0}^{\infty}\!\! dz \tfrac{r z^{3-\varepsilon}}{\left(\left(r+ R\right)^2 + z^2\right)^{4 - (1+\kappa)\varepsilon}} \left[{}_2F_1\left(\tfrac{1}{2},1 - \tfrac{(1+\kappa)\varepsilon}{4};1,\tfrac{4rR}{\left(r + R\right)^2 + z^2}\right)\right]^4\,.
\end{aligned}
\label{challenging}
\end{equation}
It is enough to give $I$ to order $\text{O}\left(\hat{\varepsilon}^0\right)$, since $\lambda_{*} \sim \hat{\varepsilon}$ and $h_{*}$ is finite. The remaining integral can be conducted using appendix B in \cite{Cuomo:2021kfm} and it evaluates to $I = 1/(16 \pi^2)$ in the limite $\hat{\varepsilon} \to 0$. 
\paragraph{Spherical integration over two angles}
\begin{equation}
\begin{aligned}
\oint \frac{d^2\tau_1 d^2\tau_2}{|x(\tau_1)-x(\tau_2)|^{2\Delta}} = \frac{(2R)^{4 - 2\Delta} \pi^2 }{1-\Delta} \,.
\end{aligned}
\end{equation}

\section{Integrating out $\hat{\psi}$}\label{integratingoutpsi}

Let's study the effective theory for $\phi$ when $\hat{\psi}$ is integrated out in a theory described by the action \eqref{nonlocaldefectaction}. The effective action thus obtained yields results identical to those derived in the two-field picture, but it is nevertheless instructive to show that the relevant functional integrals can be conducted in some simple cases. Furthermore, having noticed that in $d=4$ the dimension $\Delta_\psi \to 0$ for $a=p$, one might want to check whether or not the aforementioned action reduces to that of a localized magnetic field for $a = p = 1$ (resp. surface defect for $a=p = 2$) as $d \to 4$. 

Throughout this section, we'll be considering a free bulk, i.e. $\lambda\to 0$, and we will argue that in some cases where the defect GFF can be integrated out, there is no match with the localized magnetic field and the surface defect when $d=4-\hat{\varepsilon}$. 

Let's set $\sigma = d - 2\Delta_\phi$ and $\tau = p - 2\Delta_\psi$. To avoid conflicts and use lighter notation, we also drop the $\tau$ parametrization of the defect and instead use $y$ coordinates in this section only. It turns out that for $b=1$ or $b=2$ -- which are the only cases we'll look at -- this integration can be carried out exactly. This amounts to the following rewriting of the partition function
\begin{equation}
Z = \int \mathcal{D}\phi e^{-S_0[\phi]} \int \mathcal{D} \psi e^{-S_1[\phi,\psi]} = \int \mathcal{D} \phi e^{-S_{\text{eff}}[\phi]} \,,
\end{equation}
 where $S_0[\phi]$ is the bulk action, $S_1[\phi,\psi]$ is the defect action and the effective action is defined, up to a $\phi$-independent term, by
\begin{equation}
S_{\text{eff}}[\phi] := S_0[\phi] - \log \int \mathcal{D} \psi e^{-S_1[\phi,\psi]} \,.
\end{equation}
\subsection{$b = 1$ theory} \label{b=1 effective}
To integrate out $\psi$ in the $b = 1$ case, one can start by completing the square in the defect action
\begin{equation}
S_1[\phi,\psi] = \frac{1}{2}\int_p \left[\psi \mathcal{L}_\tau \psi + g_0 \phi^{a} \psi \right] = \frac{1}{2}\int_p \left(\psi + \frac{g_0\phi^a \mathcal{L}_{-\tau}}{4}\right) \mathcal{L}_\tau \left(\psi + \frac{g_0\mathcal{L}_{-\tau} \phi^a}{4}\right) - \frac{g_0^2}{16} \int_p \phi^a \mathcal{L}_{-\tau} \phi^a\,.
\end{equation}
 A simple translation of $\psi$ makes the first term in $S_1$ $\phi$-independent, and the resulting path integral is a constant. We are hence left with
\begin{equation}
\begin{aligned}
S_{\text{eff}}[\phi] &= \frac{1}{2}\int d^dx \phi \mathcal{L}_\sigma \phi - \frac{g_0^2}{16} \int d^p y \phi^a \mathcal{L}_{-\tau} \phi^a
\\ &= \frac{1}{2}\int d^dx \phi \mathcal{L}_\sigma \phi - \frac{g_0^2}{16} \mathcal{N}_{-\tau} \int d^p y_1 d^p y_2 \frac{\phi^a(y_1) \phi^a(y_2)}{|y_1-y_2|^{p-\tau}}\,.
\end{aligned}
\end{equation}
Taking $d \to 4$ in the $a=p$ case yields to lowest nontrivial order\,\footnote{We ignore the normalization factor in front of $g_0$, since the coupling can be redefined to absorb it as we attempt to map to the localized magnetic field or the surface defect.}
\begin{equation}
\begin{aligned}
S_{\text{eff}}[\phi] &= \frac{1}{2}\int d^dx \phi \mathcal{L}_\sigma \phi - \left(g_0\int d^py \phi^p(y)\right)^2\,,
\end{aligned}
\end{equation}
since $\Delta_\psi \to 0$. This action clearly can't be mapped to the localized magnetic field, nor to the surface defect by a simple redefinition of $g_0$.
\subsection{$b=2$ theory} \label{b = 2 effective}
In the $b = 2$  case, integrating out $\psi$ amounts to evaluating a Gaussian integral.
\begin{equation}
S_1[\phi] = \frac{1}{2}\int dx \psi \left(\mathcal{L}_\tau + g_0 \phi^a\right) \psi  \,.
\end{equation}
 This looks like a GFF, $g \phi^a$ generating a mass-like term for $\psi$. We need to calculate the following functional integral to integrate this field out
\begin{equation}
\int \mathcal{D} \psi e^{-S_1} = C \det \left(\mathcal{L}_\tau + g_0\phi^a\right)^{-1/2} = C e^{-\int d^px L_{\text{eff}}}\,,
\end{equation}
 where $C$ is some normalization constant that isn't important and we introduce an effective Lagrangian. Computing the effective action amounts to using the following identity for an operator $\hat{A}$
\begin{equation}
\log \det \hat{A} = \Tr \log \hat{A}\,.
\end{equation}
Hence the difficult task of calculating a determinant can be recast into a trace calculation. The following equalities hold modulo a $\phi$-independent term
\begin{equation}
\begin{aligned}
\Tr \log \left(\mathcal{L}_\tau + g\phi^a\right) &= \int \frac{d^pk}{(2\pi)^p} \left \langle k \left| \log \left(\mathcal{L}_\tau + g_0 \phi^a\right) \right| k \right\rangle
\\ &= \int \frac{d^pk}{(2\pi)^p} d^px e^{ikx} \left \langle x \left| \log \left(\mathcal{L}_\tau + g_0 \phi^a \right) \right| k \right\rangle
\\ &= \int \frac{d^pk}{(2\pi)^p} d^px e^{ikx} \left \langle x \left| \log \left(1 + g_0 \mathcal{L}_{\tau}^{-1} \phi^a \right) \right | k \right \rangle
\\ &= \sum_{j=1}^{+\infty} \frac{g_0^j (-1)^{j+1}}{j} \int d^px \frac{d^pk}{(2\pi)^p} e^{ikx} \left \langle x \left| \left[ \mathcal{L}_{\tau}^{-1} \phi^a \right]^j \right | k \right \rangle\,.
\end{aligned}
\end{equation}
 One can then use the following identity, easily derived using the action of the fractional Laplacian on plane waves.
\begin{equation}
\left \langle x \left| \left[ \mathcal{L}_{\tau}^{-1} \phi^a \right]^j \right | k \right \rangle = e^{-ikx} \int \prod_{i=1}^j \frac{d^pq_i}{(2\pi)^p} \hat{\phi^a}(q_i) e^{iq_ix} \frac{1}{\left|q_1 + \dots + q_i - k\right|^\tau}\,.
\end{equation}
 We finally get $L_{\text{eff}} = \sum_j L_{\text{eff}}^{(j)}$ with
\begin{equation}
\begin{aligned}
L_{\text{eff}}^{(j)} &= \frac{g_0^j (-1)^{j+1}}{2j} \int \frac{d^pk}{(2\pi)^p} \prod_{i=1}^j \frac{d^pq_i}{(2\pi)^p} \hat{\phi^a}(q_i) e^{iq_ix} \frac{1}{\left|q_1 + \dots + q_i - k\right|^\tau}
\\ &= \frac{g_0^j (-1)^{j+1}}{2j} \phi(x)^a \int \frac{d^pk}{(2\pi)^p} \frac{1}{|k|^{\tau}} \prod_{i=2}^j \frac{d^pq_i}{(2\pi)^p} \hat{\phi^a}(q_i) e^{iq_ix} \frac{1}{\left|q_2 + \dots + q_i - k\right|^\tau}
\\ &= \frac{g_0^j (-1)^{j+1}}{2j} \phi(x)^a \int \frac{d^pk}{(2\pi)^p} \frac{1}{|k|^{\tau}} \prod_{i=2}^j \frac{d^pq_i}{(2\pi)^p} d^px_i \phi(x_i)^a e^{iq_i(x-x_i)} \frac{1}{\left|q_2 + \dots + q_i - k\right|^\tau}\,.
\end{aligned}
\end{equation}
Reorganizing the complex exponentials allows one to rewrite this using known propagators, yielding
\begin{equation}
S_{\text{eff}} = \sum_{j=1}^{+\infty} \frac{(-1)^{j+1} \left(\mathcal{N}_\psi^2\right)^j}{2j} g_0^j \int \prod_{i=1}^j d^px_i \frac{\phi(x_i)^a}{|x_{i+1} - x_i|^{p-\tau}}\,.
\end{equation}
 Setting $a=p = 1$ or $2$ and taking $d \to 4$, the effective action becomes, to lowest nontrivial order
\begin{equation}
S_{\text{eff}}[\phi] = \frac{1}{2}\int d^dx \phi \mathcal{L}_\sigma \phi + \frac{1}{2} \log \left(1 + g_0 \int d^py \phi(y)^p\right)\,.
\end{equation}
This action seems a lot closer to the localized magnetic field and the surface defect. As we have seen, $g$ renormalizes in a few of the interesting cases, and the fixed point is infinitesimal. Hence, expanding the logarithm leads to
\begin{equation}
S_{\text{eff}} = \frac{1}{2}\int d^dx \phi \mathcal{L}_\sigma \phi + \frac{g_0}{2} \int d^py\phi(y)^p + \dots
\end{equation}
Therefore, at the lowest orders, the theory does resemble the line or surface defects we studied. Notwithstanding, their behaviors diverge from one another as we include higher loop orders. One observable which exemplifies this is the $g$-function for the $a=p=1$ theory, which is readily accessible from the effective theory
\begin{equation}
\begin{aligned}
g = \sum_{k=0}^{\infty} \frac{\left(4k\right)!}{\left(2k\right)!k!} \left[g_0^2 2^{-3 - 2\Delta_\phi} R^{2-2\Delta_\phi} \pi \frac{\Gamma\left(\frac{1}{2} - \Delta_\phi\right) \sqrt{\pi}}{ \Gamma\left(1 - \Delta_\phi\right)}\right]^{k}\,.
\end{aligned}
\end{equation}
 The above expression cannot directly be mapped to \eqref{free g} by a redefinition of $g_0$. This is essentially because of the coefficients $(4k)!/(2k)!$ making it depart from the series of an exponential. Incidentally, these coefficients also translate the different combinatorics in the two theories on the line. Indeed, when $\Delta_\psi \to 0$, some diagrams which were previously distinct become the same, and contribute non-trivially to the combinatorics.

\newpage
\bibliographystyle{nb}
\bibliography{references}

\end{document}